\documentclass[useAMS]{mn2e}
\usepackage{graphicx}
\usepackage{epsfig}
\usepackage{amssymb}
\usepackage{lscape}
\usepackage{ulem}
\usepackage{txfonts}

\def\sigsfr{$\Sigma_{\rm SFR}$}

\def\sigmass{$\Sigma_{\star}$}
\def\dsfr{$\Delta$SFR}
\def\dsfrs{$\Delta \Sigma_{\rm SFR}$}
\def\kms{km~s$^{-1}$}

\def\c2s{C\,{\sc ii}$^{\star}$}

\def\Ha{H$\alpha$}

\def\OIII{O{\sc iii}}

\def\NII{N{\sc ii}}

\def\HII{H{\sc ii}}

\def\dsfr{$\Delta$SFR}

\title[Star formation profiles with MaNGA] {Star formation is boosted (and quenched) from the inside out: radial star formation profiles from MaNGA.}

\author[Ellison et al.] {Sara L. Ellison$^1$, Sebastian F. S\'{a}nchez$^2$, Hector Ibarra-Medel$^2$, Braulio Antonio$^{1,2}$, 
\newauthor J. Trevor Mendel${^3}$,  Jorge Barrera-Ballesteros$^4$\\
$^1$ Department of Physics \& Astronomy, University of Victoria, Finnerty Road, Victoria, British Columbia, 
V8P 1A1, Canada.\\
$^2$ Instituto de Astronomıa, Universidad Nacional Autonoma de Mexico, A. P. 70-264, C.P. 04510, Mexico, 
D.F., Mexico.\\
$^3$ Max-Planck-Institut fur Extraterrestrische Physik, Giessenbachstrasse, D-85748 Garching, Germany.\\
$^4$  Department of Physics \& Astronomy, Johns Hopkins University, Bloomberg Center, 3400 N. 
Charles St., Baltimore, MD 21218, USA.
}
\begin{document}

\maketitle

\begin{abstract}
  The tight correlation between total galaxy stellar mass and star formation rate (SFR)
  has become known as the star forming main sequence.  Using $\sim$ 487,000 spaxels from 
 galaxies observed as part of the
  Sloan Digital Sky Survey Mapping Galaxies at Apache Point Observatory (MaNGA)
  survey, we confirm previous results that a correlation also exists
  between the surface densities of star formation (\sigsfr) and 
stellar mass (\sigmass) on kpc scales, representing a `resolved' main sequence.  
Using a new metric (\dsfrs), which measures the relative enhancement or deficit of
star formation on a spaxel-by-spaxel basis relative to the resolved main sequence, we investigate
  the SFR profiles of 864 galaxies as a function of their position relative to the global
  star forming main sequence (\dsfr).    For galaxies above the global main sequence
(positive \dsfr) \dsfrs\ is elevated throughout the galaxy, but the greatest enhancement
in star formation occurs at small
  radii ($<$ 3 kpc, or 0.5 $R_e$).  Moreover, galaxies that are at least a factor of three above
  the main sequence show diluted gas phase metallicities out to 2 $R_e$, indicative of metal-poor 
gas inflows accompanying the starbursts.  For quiescent/passive galaxies that lie at least a
  factor of 10 below
the star forming main sequence there is an analogous deficit of star formation
throughout the galaxy with the lowest values of \dsfrs\ in the central 3 kpc.
Our results are in qualitative 
agreement with the `compaction' scenario in which a central starburst leads to
mass growth in the bulge and may ultimately precede galactic quenching from the inside-out.

\end{abstract}


\section{Introduction}

One of the key contributions of large galaxy surveys has been to establish
global scaling relations between galaxy properties.  One such correlation
has become known as the star forming main sequence (hereafter,
simply `main sequence'\footnote{We will sometimes refer to the main
sequence relationship between galaxy total stellar mass and total star
formation rate as the `global'
main sequence, in order to distinguish it from the `resolved' main
sequence that is also studied in this work.}) - a tight relation between a galaxy's star formation
rate (SFR) and its total stellar mass (Brinchmann et a. 2004; Salim et al. 2007;
Renzini \& Peng 2015).  This relationship between SFR and M$_{\star}$ 
exists out to high redshifts, increasing its normalization to higher
values at earlier epochs such that SFRs at a fixed stellar mass
are higher by a factor of $\sim$ 20 by $z \sim 2$ (e.g. Noeske et al. 2007;
Daddi et al. 2007; Whitaker et al. 2012; Speagle et al. 2014; 
Schreiber et al. 2015; Barro et al. 2017).

Galaxies above/below the main sequence at any given epoch have, by definition, 
SFRs that are elevated or suppressed relative to the `norm' for their stellar mass.  Various
processes have been associated with the modulation of star formation rates,
including internal structure (such as bars and bulges, e.g. Ellison
et al. 2011; Wuyts et al. 2011b; Wang et al. 2012; Mendel et al. 2013; Bluck et al. 2014), cold 
gas fraction (Saintonge et al. 2012, 2016;  Tacconi et al. 2013, 2017;  Sargent et al. 2014; 
Genzel et al. 2015; Violino et al. 2017) interactions with
other galaxies (Ellison et al. 2008a, 2013; Scudder et al. 2012; Patton et al. 2013;
Scott \& Kaviraj 2014; Stierwalt et al. 2015; Willett et al. 2015) and the presence of an AGN
(Shimizu et al. 2015; Cowley et al 2016; Ellison et al. 2016; Azadi et al. 2017;
S\'{a}nchez et al. 2017b).
The fraction of galaxies that inhabit the main sequence is also a function of both
local and large scale environment (Peng et al. 2012; Woo et al. 2013; Knobel et al.
2015), with lower star formation rates in higher density environments 
(Lewis et al 2002;  Gomez et al. 2003).  These various processes for modulating
star formation may be expected to impact the
internal profiles of star formation in different ways.  For example, positive
feedback from an AGN, which frequently manifest evidence for central winds
(Crenshaw et al. 2010; Harrison et al. 2014; McElroy et al. 2015; Woo et al. 2016, 2017; 
Concas et al. 2017) 
could suppress the star formation preferentially in the inner galactic regions
(e.g. Cano D\'{i}az et al. 2012; Carniani et al. 2016).
Conversely, both bars and galaxy-galaxy interactions can funnel gas
inwards (Martel et al. 2013; Moreno et al. 2015) leading to central SFR enhancements.
Other processes, such as simple gas exhaustion, or stochastic bursts
of star formation due to instabilities in gas rich disks, 
may lead to a more uniform SFR suppression/enhancement.

In recognition that, in practice, a variety of mechanisms
can trigger star formation, a more generalized paradigm has 
recently emerged from simulations and observations alike.  
In this model, varying processes such as mergers (both minor and 
major), secular disk instabilities and  streams lead to high
gas densities and centrally concentrated star formation (e.g. 
Dekel \& Burkert 2014;
Zolotov et al. 2015; Tacchella et al. 2016a,b).  Due to
the characteristic build-up of central stellar mass density
that follows the starburst, this process has been termed
`galaxy compaction'.
Galaxies can oscillate around the global star forming main sequence
as a result of successive compaction and gas depletion events
(e.g. Fig. 11 in Tacchella et al. 2016a).  Empirical support for
this compaction process comes from a tight observed correlation
between the central surface density of stellar mass (\sigmass) and
total stellar mass, wherein quiescent galaxies are offset to
higher central \sigmass\ at fixed M$_{\star}$ (Fang et al. 2013; Tacchella
et al. 2015a), as well as the existence of a population of star forming galaxies
with the same central \sigmass\ as quiescent galaxies 
(Barro et al. 2013, 2014, 2017).  These `compact star-forming'
galaxies have been proposed as the possible pre-cursors of fully
quenched galaxies.

A clear testable prediction of the compaction model is that elevation 
above the main sequence is driven by central star formation,
and that quenching proceeds with the same radial directionality. 
There is now considerable empirical evidence, from a variety of datasets,
redshifts and observational techniques that support inside out
quenching (e.g  Gonz\'{a}lez Delgado et al. 2016; Nelson et al.
2016;  Belfiore et al. 2017a, Tacchella et al. 2015a, 2017;
S\'{a}nchez et al. 2017b; Morselli et al., in prep).  
There is likewise support for the importance
of the central regions in building the galaxy's stellar mass and
inside out growth (Nelson et al. 2012; P\'{e}rez et al. 2013; Gonz\'{a}lez
Delgado et al. 2014, 2015; Morselli et al. 2017; Lian et al. 2017).  However, simultaneous
assessments of the radial dependence of star formation both above
and below the main sequence using the same dataset and homogeneous
analysis are rare. From a study of $\sim$ 3000 galaxies at
$z \sim 1$, Nelson et al. (2016) found that, in general, star formation 
was uniformly suppressed/enhanced in galaxies below/above the
main sequence.  Only in the highest mass galaxies in their sample
were centrally enhanced trends evident.  Simulations have shown that
these observational results can be reproduced by bursty star formation 
histories (Orr et al. 2017).  Evidence
for centrally suppressed specific SFRs in galaxies below the main 
sequence at $z \sim 1$, as predicted by compaction,
have been found by Morselli et al. (in prep),
but this sample lacks the statistics to study true starbursts.  Tacchella
et al. (2017) present tantalizing evidence for centrally driven
radial changes in galaxies above and below the main sequence
at $z \sim 2$, but with a sample of only 10 galaxies, this
remains tentative.

Large integral field unit (IFU) galaxy surveys such as the Calar Alto 
Legacy Integral Field Area (CALIFA, S\'{a}nchez et al. 2012), 
Sydney-Australian-Astronomical-Observatory Multi-object 
Integral-Field Spectrograph (SAMI, Croom et al. 2012; Allen et al. 2015) 
and the Mapping Nearby Galaxies and
Apache Point Observatory (MaNGA, Bundy et al. 2015) have the potential to
revolutionize our ability to map star formation in galaxies and test
radial trends in galaxy evolution.
In the current work, we seek to use a large IFU sample of galaxies from
the MaNGA survey to address the question
of where within a galaxy the star formation is being regulated, both above and below
the global star forming main sequence.  
The paper is laid out as follows.  In Section 2 we describe our sample
selection from the MaNGA galaxy survey, as well as the definition of the SFR offset metric used
to quantify the position of a given galaxy relative to the main sequence.
In Section 3 we describe the analysis pipeline applied to the IFU data cubes,
measurement of spaxel properties, the resolved main sequence in
MaNGA star forming spaxels and new metrics developed to quantify
the relative enhancement/suppression of star formation and metallicity on a spaxel
by spaxel basis.  In Section 4 we present the main results of our
study -- relative star formation profiles as a function of offset
from the global main sequence.  We discuss our results in Section 5
and summarize in Section 6.  We adopt a cosmology in which
H$_0$=70 km/s/Mpc, $\Omega_M$=0.3, $\Omega_\Lambda$=0.7.

\section{Sample selection}

\subsection{MaNGA parent sample}\label{sample_sec}

In this work we use the galaxies available in the Sloan Digital Sky Survey (SDSS)
Data Release 13 (DR13; Albareti et al. 2017) observed as part
of the MaNGA survey.  The MaNGA survey is one of three projects within SDSS-IV
that will ultimately target 10,000 galaxies evenly sampled above a
stellar mass log (M$_{\star}$/M$_{\odot}) \sim 9$ (Bundy et al. 2015).
All of the MaNGA targets are selected from the SDSS main galaxy sample,
offering the benefit of previously determined global properties such
as metallicities, morphologies, SFRs and stellar masses (e.g. Kauffmann et al. 2003b;
Brinchmann et al. 2004; Salim et al. 2007; Simard et al. 2011; Mendel et al. 2014).
By bundling together the individual 2$\arcsec$ fibres of the twin Baryon 
Oscillation Spectroscopic Survey (BOSS) spectrographs into hexagonal 
IFUs, and employing a dithering strategy to fill in the gaps 
between fibres, a continuous spectral map of the galaxy can be obtained 
(Law et al. 2015).  The IFUs vary in diameter from 12$\arcsec$ (19 fibres) to 32$\arcsec$ 
(127 fibres) and are selected to cover any given galaxy out to
1.5 effective radii for 2/3 of the sample.  The remainder of the
sample is selected at slightly higher redshifts in order to achieve
coverage out to 2.5 effective radii.

A query of all the publically available data in
the MaNGA DR13 sample yields 1390 datacubes, including a minority of
duplicate observations of the same target galaxy.  In this work, we
will make use of several extant catalogs of derived galaxy properties,
primarily based on the SDSS DR7.  These include measurements of stellar
mass (Kauffmann et al. 2003b), star formation rates (Brinchmann et al. 2004;
Salim et al. 2007), galaxy half light (effective) radius in the $r$-band
($R_e$, Simard et al. 2011), bulge fractions measured in the $r$-band
(Simard et al. 2011), bulge fractions as determined from the 
stellar mass (Mendel et al. 2014) and AGN classification
(Kauffmann et al. 2003a). We therefore require that to be included our sample,
a galaxy in the DR13 must also be included in all of these aforementioned data 
catalogs, for which we require a positional match within 2$\arcsec$.
There are 1157 unique galaxies in the DR13 that are matched to the DR7 catalogs
within this tolerance.

\subsection{Star forming galaxies}\label{sf_sec}

\begin{figure}
	\includegraphics[width=\columnwidth]{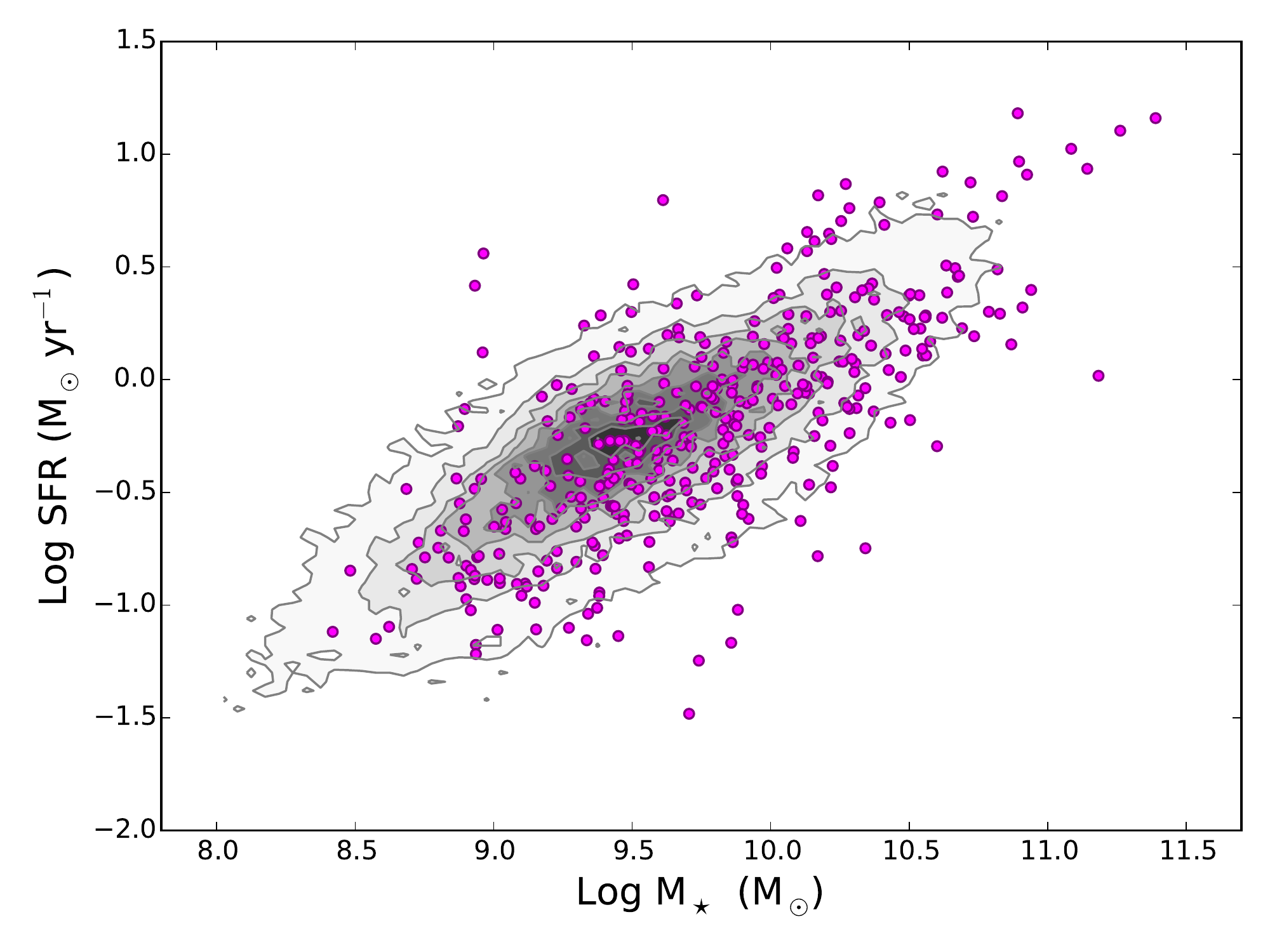}
        \caption{The star forming main sequence as defined by $\sim$ 65,000 z$<$0.06
star forming galaxies from
        the SDSS DR7 (grey contours).  Magenta points show the positions of 394 
star forming MaNGA  galaxies selected with the same criteria.  Whilst all 394
galaxies are shown here for reference, two galaxies with 
\dsfr\ $<-1.0$ are transferred from the final star forming sample into the passive
sample. }
    \label{MS}
\end{figure}

Our study will distinguish star-forming and quenched (passive) galaxies.
Star forming galaxies were selected
according to the following criteria:  a stellar mass and total SFR must be available
from the MPA/JHU catalog (Kauffmann et al. 2003b; Brinchmann et al. 2004; Salim et al.
2007) and the galaxy must be classified as star-forming according to the emission line
ratio criteria of Kauffmann et al. (2003a), with a minimum S/N=3 required
for all relevant emission lines in the DR7 spectrum.  There are $\sim$156,000 galaxies 
in the DR7 that fulfill the criteria required for our star forming sample, of which 394 
are part of the MaNGA DR13.  In Fig. \ref{MS} we show the distribution of SFR
and M$_{\star}$ of the 394 star forming MaNGA galaxies as magenta points.  For
reference, the DR7 distribution (restricted to galaxies with $z<0.06$ for display purposes,
in order to represent the dominant redshift range of the MaNGA sample) is shown 
in grey contours.

We note that total SFRs and stellar mass estimates can also be computed for the
MaNGA sample by integrating across all the spaxels in the IFU.  However, in this work 
(see Section \ref{dsfr_sec}) we will be computing galaxy offsets from the star 
forming main sequence (\dsfr), by comparing the SFRs of MaNGA galaxies to the full 
sample of SDSS DR7 galaxies at fixed M$_{\star}$, redshift and local
galaxy density.  In order to have consistent galaxy stellar mass and SFR
measurements for the \dsfr\ calculation, we adopt the DR7 measurements of
these values from the MPA/JHU catalogs.  Using only the MaNGA galaxies to
define the control samples leads to a poor statisical matching (very few matches
at the extremes of stellar mass and density).  However,
we have checked that for the MaNGA galaxies the stellar mass and SFR
values from the MPA/JHU catalog correlate with the integrated
MaNGA values.  The mean difference between the MPA/JHU and integrated MaNGA
values is 0.0007 dex for stellar mass and 0.03 dex for SFR, with scatter of
$\sim$ 0.3 and 0.4 dex respectively (consistent with comparisons in other
papers, e.g. Spindler et al. 2017), see Appendix A.  

\subsection{Star formation rate offsets from the main sequence}\label{dsfr_sec}

\begin{figure}
	\includegraphics[width=\columnwidth]{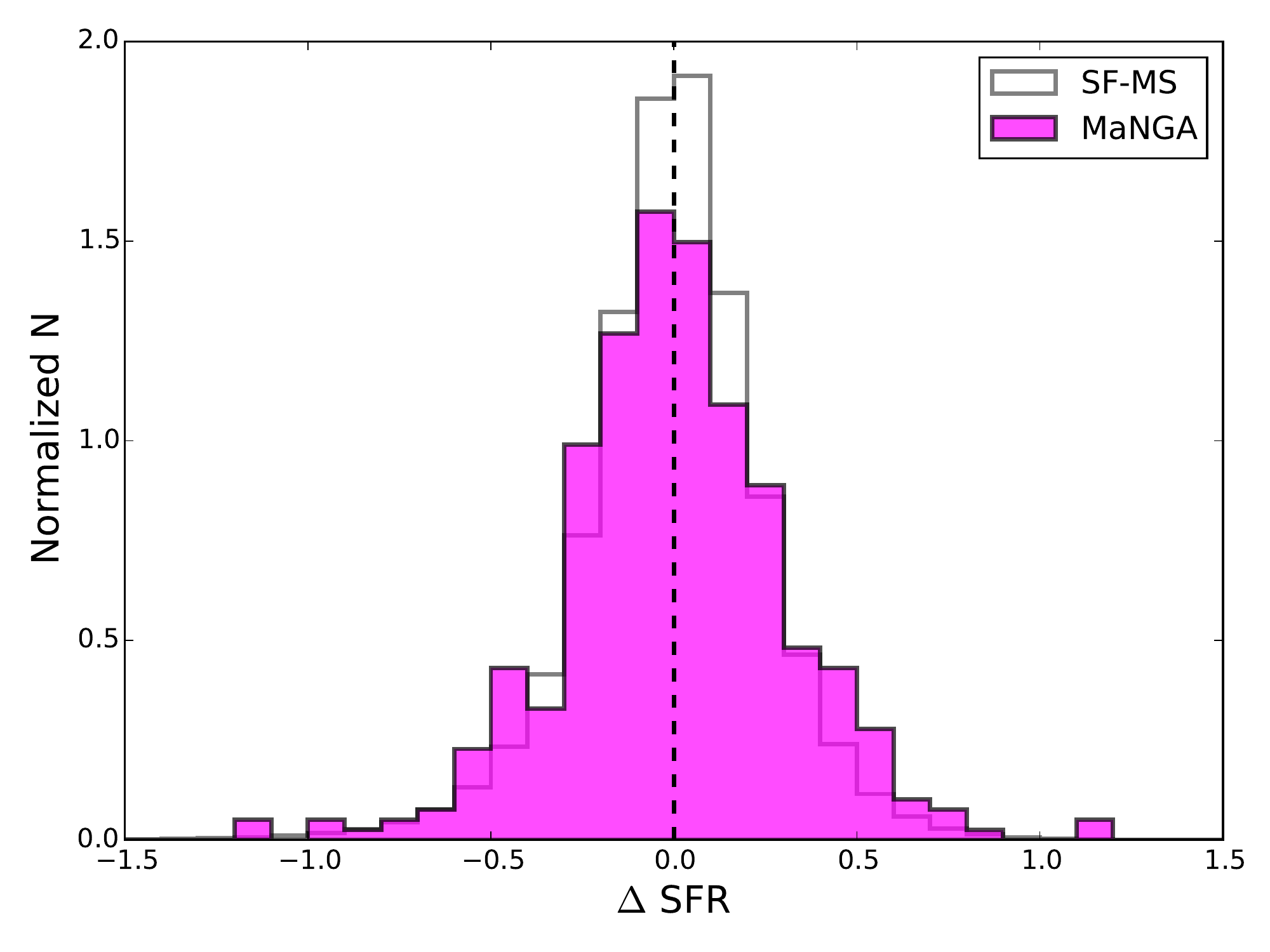}
        \caption{Normalized histogram of \dsfr\ for the DR7 and MaNGA samples of star forming galaxies presented in Fig. \ref{MS}. The white histogram shows the distribution of \dsfr\ for the full DR7 star-forming sample
          ($\sim$ 156,000 galaxies) that define the global main sequence.
          The magenta histogram shows the distribution of \dsfr\ for the 394 star-forming
        galaxies in MaNGA used in this work. Whilst all 394
galaxies are shown here for reference, two galaxies with 
\dsfr\ $<-1.0$ are transferred from the final star forming sample into the passive
sample.}
    \label{ms_dsfr}
\end{figure}

The main goal of this work is to investigate the spatial profiles of galaxies
that exhibit different global levels of star formation.  It is
therefore necessary to quantify whether (and by how much) a galaxy
is forming stars at a higher or lower rate than expected, given its
other various properties.
In order to quantify how enhanced or suppressed the SFR is in a given galaxy,
relative to the `norm', we define a SFR offset (\dsfr).
Qualitatively, \dsfr\ is the difference (on a logarithmic scale) between
the observed SFR of a given galaxy and its expected SFR (defined quantitatively
below).  Hence a \dsfr\ = 1 indicates a galaxy whose SFR is elevated 
above the expected value on the main sequence by a factor of 10.
A further benefit of computing a differential measure of star formation
is that it helps to mitigate biases in sample selection and parameter
determination, such as aperture effects (e.g. Richards et al. 2016;
Duarte Puertas et al., 2017).

Since the main factor that regulates a galaxy's SFR is its stellar mass, a simple
approach would be to fit a relation to the main sequence or simply
to compare to the peak SFR at a given M$_{\star}$ (e.g. Renzini \& Peng 2015;
Magdis et al. 2016; Morselli et al. 2017).
However, other factors may also modulate the
global galactic SFR, hence it is desirable to account for additional
parameters.  Firstly, since local galaxy density has been shown to affect star formation (e.g.
Lewis et al. 2002; Gomez et al. 2003), it is desirable to match on
some environmental metric.  Following our previous work,
we adopt the quantity

\begin{equation}
\Sigma_5 = \frac{5}{\pi d_5^2},
\end{equation}

where $d_5$ is the projected distance in Mpc to the $5^{th}$ nearest
neighbour within $\pm$1000 \kms.  Normalized densities, $\delta_5$,
are computed relative to the median $\Sigma_5$ within a redshift slice
$\pm$ 0.01.

Secondly, we match on galaxy redshift.
Matching on redshift is potentially important for two reasons.  First, if any fibre based
quantity is used in the analysis, aperture corrections will evolve
significantly over the redshift range of the DR7 sample.  This
should not be a significant effect in our analysis of SFRs, since we
use aperture corrected values (e.g. Brinchmann et al. 2004;
Salim et al. 2007).  The second reason to match in redshift
is to account for a true
evolution of sample properties.  As noted in the Introduction,
it is well known that the star forming main sequence increases its
normalization towards higher redshifts (e.g. Whitaker et el.
2012; Speagle et al. 2014).  Even within the relatively narrow
redshift range of the SDSS DR7 (where the maximum redshift is $z \sim
0.2$), the  main sequence normalization changes by approximately 0.2 dex.  
 
In order to compute the \dsfr\ of a given galaxy (SFR$_{\rm gal}$),
we construct a sample of control star forming galaxies
that are matched in stellar mass, redshift and local galaxy density
(environment) that are drawn from the DR7 parent sample of
$\sim$ 156,000 star-forming galaxies described above.
The baseline tolerance used for matching is 0.1 dex
in stellar mass, 0.005 in redshift and 0.1 dex in $\delta_5$.
We require at least five comparison galaxies in the matched sample;
if this is not achieved then the mass, redshift and local density tolerances
are grown in further increments of 0.1 dex, 0.005 and 0.1 dex respectively,
until the minimum size criterion of five matched controls is achieved.  In practice, 95 per cent 
of galaxies are successfully matched to at least five controls without
the need to grow the tolerances.  The remaining five percent require only
one `grow' in order to reach the requirement of five matched controls. 
In general, the number of matched
controls far exceeds the minimum requirement of five, with an average of
110 matches per galaxy.  

The SFR of the control star 
forming sample (SFR$_{\rm control}$) is taken as the median
of the aperture-corrected `total' SFRs determined from the SDSS spectra 
(Brinchmann et al. 2004; Salim et al. 2007).The SFR offset is then defined as:

\begin{equation}\label{eqn_dsfr}
  \Delta SFR = \log SFR_{\rm gal} - \log SFR_{\rm control}.
\end{equation}

In Fig. \ref{ms_dsfr} we show the distribution of \dsfr\ for the galaxies
in the star forming MaNGA and DR7 galaxies presented in Fig. \ref{MS}.
By construction, the DR7 sample is symmetric around zero.    The MaNGA
sample is also broadly symmetric, spanning a wide range of \dsfr, including galaxies that exhibit
SFRs up to 10 times above or below their matched control samples.
In order to cleanly distinguish the star forming sample of MaNGA
galaxies (Sec. \ref{sf_sec}) from the passive MaNGA galaxies
(Sec. \ref{pass_sec}) we impose a cut on the \dsfr\ of the
former sample, requiring that \dsfr\ $> -1.0$.  This excludes
two galaxies from the original 394 in the MaNGA star-forming sample;
these two galaxies are instead considered as part of the passive
galaxy sample.

\subsection{Passive galaxies}\label{pass_sec}

\begin{figure}
	\includegraphics[width=\columnwidth]{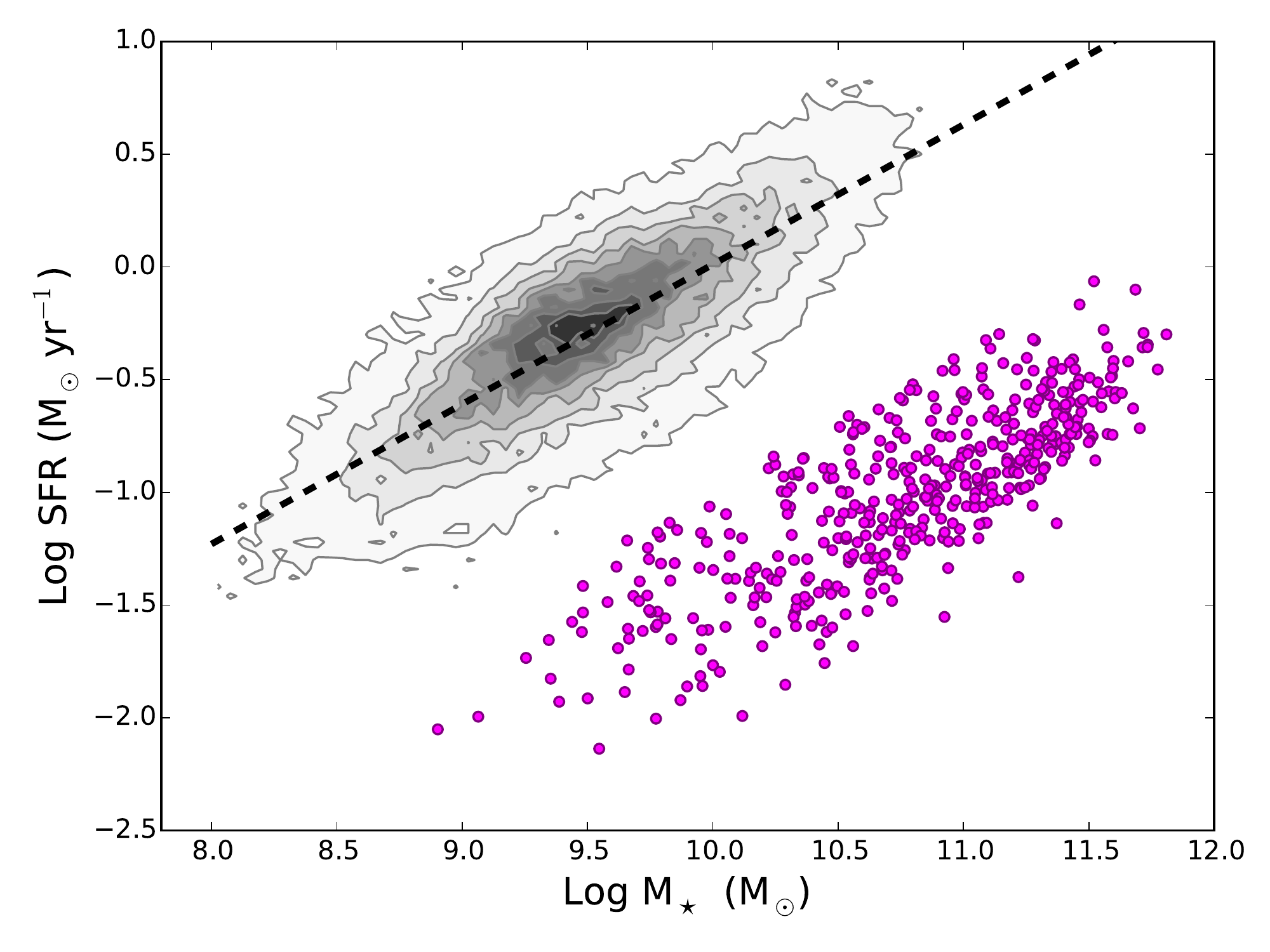}
        \caption{The star forming main sequence as defined by $\sim$ 65,000 $z<0.06$ 
star forming galaxies from
        the SDSS DR7 (grey contours).  Magenta points show the positions of 470 MaNGA 
galaxies whose SFRs are at least a factor of 10 below the best fit (dashed line) through the
main sequence and hence defined as passive.   }
    \label{MS_passive}
\end{figure}

In addition to the star forming galaxy sample, we select galaxies
from the MaNGA DR13 parent sample that are classified as 
`passive' (sometimes also referred to as `quenched'; indeed, we will
use the two terms interchangeably in this work).  Passive
galaxies are no longer actively forming stars and fall significantly
below the main sequence.  Following Bluck et al. (2016)
we defined passive galaxies as those which lie at least one
dex below the main sequence, i.e. have SFRs at least 10 times lower
for their stellar mass.  In the parlance of the previous
sub-section, passive galaxies have
\dsfr $< -1.0$.  However, since the SFRs of passive galaxies (which
are determined from a measurement of the 4000 \AA\ break, Brinchmann
et al. 2004) have large uncertainties (e.g. Rosario et al. 2016) we
do not use the \dsfr\ metric for their selection.  Instead, we
determine the best fit to the star forming main sequence and
then identify passive MaNGA galaxies as those whose SFRs are at least
a factor of 10 below the fit, without further quantifying how
far below this threshold they lie.  In this way, we acknowledge that
the exact SFRs of the passive galaxies are uncertain, but
simply use their approximate location relative to the main
sequence to identify their passive nature.
As described above, the main
sequence can vary even within the redshift range of the DR7 sample.
We therefore use the 64,505 star-forming galaxies with $z<0.06$
(typical of the MaNGA sample) from
the SDSS DR7 to fit the main sequence, as shown by the grey contours 
in Fig. \ref{MS_passive} (note the different axis ranges compared with 
Fig. \ref{MS}).  The magenta points in  Fig. \ref{MS_passive}
show the 470 MaNGA galaxies that are classified as passive.

\section{Data cube processing and spaxel quantities}

\subsection{Spectral fitting with \textsc{pipe3d}}\label{pipe3d_sec}

{\sc Pipe3D} is a software package designed to fit the
stellar continuum and measure the emission line fluxes
of IFU data (S\'{a}nchez et al. 2016a, 2016b).  The {\sc Pipe3D}
code is used in this work to determine all of the spaxel
emission line fluxes, surface densities of star formation
(\sigsfr) and stellar mass (\sigmass).

The current implementation of {\sc Pipe3D} adopts the GSD156 library
of simple stellar populations (SSPs, Cid-Fernandes et al. 2013), that
comprises 156 templates covering 39 stellar ages (from 1 Myr to
14.1 Gyr), and 4 metallicities (Z/Z$_{\odot}$=0.2, 0.4, 1, and 1.5). These
templates have been extensively used within the CALIFA collaboration
(e.g. P\'{e}rez et al. 2013; Gonz\'{a}lez Delgado et al. 2014), and for other surveys
(e.g. Ibarra-Medel et al. 2016). Details of the fitting procedure, dust
attenuation curve for the stellar population, and uncertainties on the
processing are given in S\'{a}nchez et al. (2016a, 2016b), but we provide a brief
summary here.

A spatial binning is first performed in order to reach a
S/N of 50 measured in the range 5590 -- 5680 \AA
accross the entire field of view (FoV) for each datacube. A
stellar population fit of the coadded spectra within each spatial bin
is then computed. The fitting procedure involves two steps: first, the
stellar velocity and velocity dispersion are derived, together with
the average dust attenuation affecting the stellar populations
(A$_{V,ssp}$).    In the second step,
a multi-SSP linear fitting is performed, using
the library described before and adopting the kinematics and dust
attenuation derived in the first step. This second step is repeated
including perturbations of the original spectrum within its errors;
this Monte-Carlo procedure provides the best coefficients of the
linear fitting and their errors, which are propagated for any further
parameter derived for the stellar populations.

We estimate the stellar population model for each spaxel by re-scaling
the best fit model within each spatial bin to the continuum flux
intensity in the corresponding spaxel, following
Cid-Fernandes et al. (2013) and S\'{a}nchez et al. (2016a). These
model spectra are then subtracted from the original cube to create a
gas pure cube comprising only the ionised gas emission lines (and the
noise). Individual emission line fluxes were then measured spaxel by
spaxel using both a single Gaussian model for each emission line and
spectrum, and a weighted momentum analysis, as described in
S\'{a}nchez et al. (2016b). Dust extinction is computed on a
spaxel-by-spaxel basis using the H$\alpha$/H$\beta$ ratio. 
An intrinsic value of 2.86 is assumed for this ratio.  
Corrections for extinction are made to emission line fluxes in each spaxel 
by assuming a Galactic extinction law
following Cardelli, Clayton \& Mathis (1989), with R$_{\rm V}$=3.1.

The star formation rate surface densities
were derived using all the \Ha\ intensities for all the
spaxels with detected ionized gas. The intensities are transformed to
luminosities (using the adopted cosmology) and corrected for dust
attenuation as described above. Finally we apply the
Kennicutt (1998) calibration to obtain the spatially-resolved
distribution of the SFR surface density.  Initially, SFRs are
computed for all the spaxels
irrespective of the origin of the ionization. By doing so, we take
into account the point spread function (PSF) wings in the star-forming regions, that may
present equivalent widths (EWs) below the cut applied in
S\'{a}nchez et al. (2017a) and Cano-D\'{i}az et al. (2016). However, we describe
below that only star-forming spaxels are used in the science
analysis.

\subsection{The resolved star forming main sequence}\label{res_MS_sec}

\begin{figure}
	\includegraphics[width=\columnwidth]{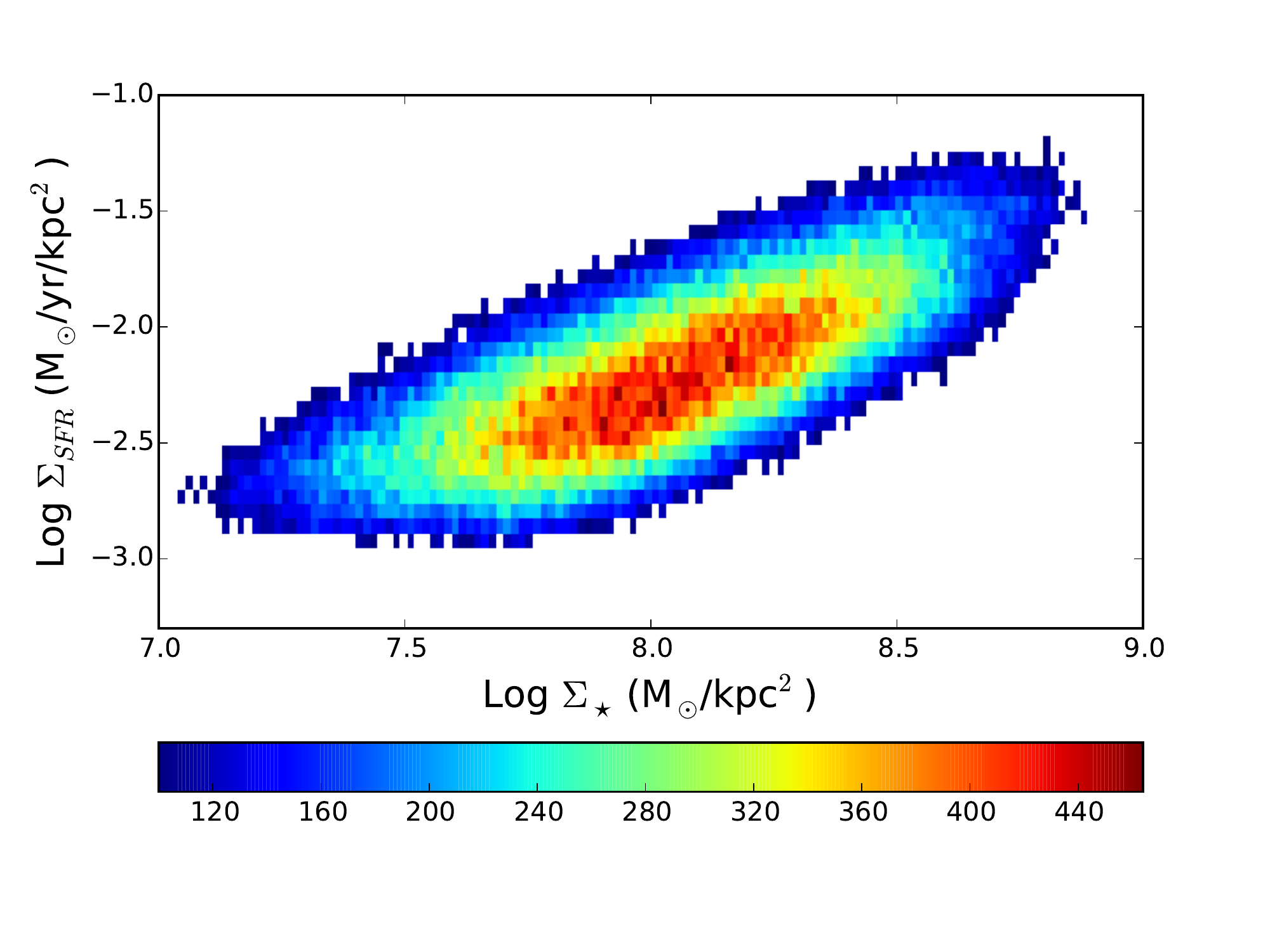}
        \caption{The local (`resolved') star forming main sequence for $\sim$487,000
          star-forming spaxels in the MaNGA DR13 datacubes.
          The colour bar indicates the number of spaxels in 
each bin. The minimum \sigsfr\ is set by our definition of
the star forming spaxel sample.  }
    \label{manga_sfms}
\end{figure}

Multi-wavelength imaging and IFU surveys alike 
have revealed that the relationship between SFR and stellar mass that
is known on global scales also exists on local (kpc) scales.  
This so-called `resolved' star forming main sequence manifests
as a tight correlation between \sigsfr\ and \sigmass, with a slope
that matches that of the global star-forming main sequence, both
locally  (e.g. S\'{a}nchez et al. 2013;
Cano-D\'{i}az et al. 2016; Gonz\'{a}lez-D\'{i}az et al. 2016;
Abdurro'uf \& Akiyama 2017; Maragkoudakis et al. 2017; Hsieh et al. 2017)
and at high redshift (e.g. Wuyts et al 2013; Magdis et al. 2016).

In Fig. \ref{manga_sfms}
we show the `resolved' star forming main sequence derived from star-forming
spaxels in the MaNGA DR13 datacubes.  Spaxels are considered as star-forming if they
have a measured value of $\Sigma_{\star}$ and \sigsfr, are designated as
star-forming by the Kauffmann et al. (2003a) emission line criteria and
have S/N$>$3 in all 4 diagnostic emission lines used therein. 
Out of $\sim$ 2 million spaxels with \sigsfr\ measured by PIPE3D, there are
$\sim$ 487,000 star-forming spaxels in the MaNGA DR13 sample according
to the above criteria. We note that these star
forming spaxels can be taken from any galaxy, including
galaxies not classified as star-forming based on their global spectroscopy, as
long as the spaxel itself is classified as star-forming.  The S/N
criteria that we impose result in an effective \sigsfr\ sensitivity
down to log \sigsfr $\sim -3$ (Fig \ref{manga_sfms}).  We
have experimented with both relaxing and tightening the spaxel
S/N requirement and although it does impact the effective \sigsfr\
threshold, it does not qualitatively alter the conclusions of this
work.  Further discussion of selection biases is presented in
Sec. \ref{bias_sec}.

\begin{figure}
	\includegraphics[width=\columnwidth]{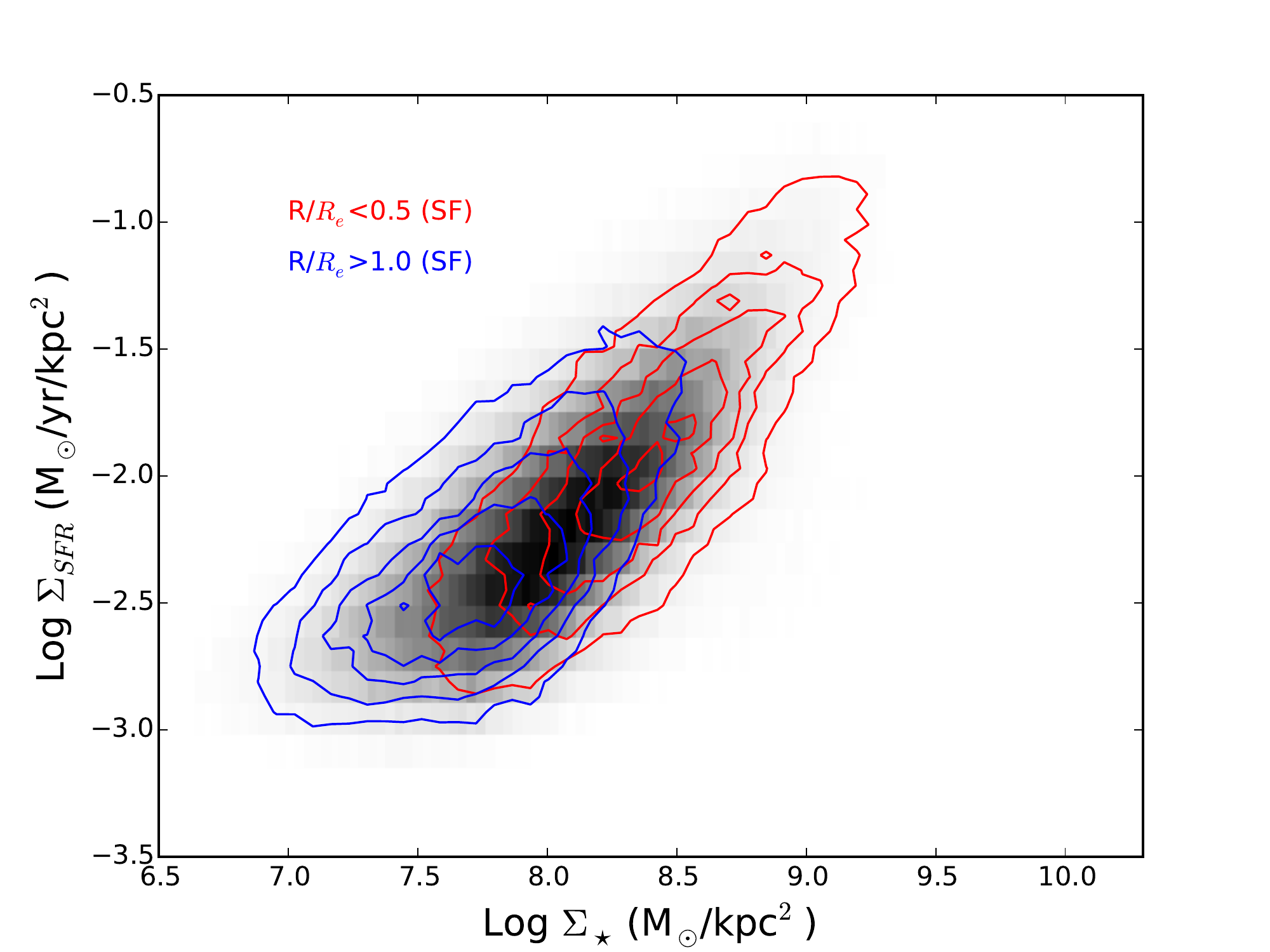}
	\includegraphics[width=\columnwidth]{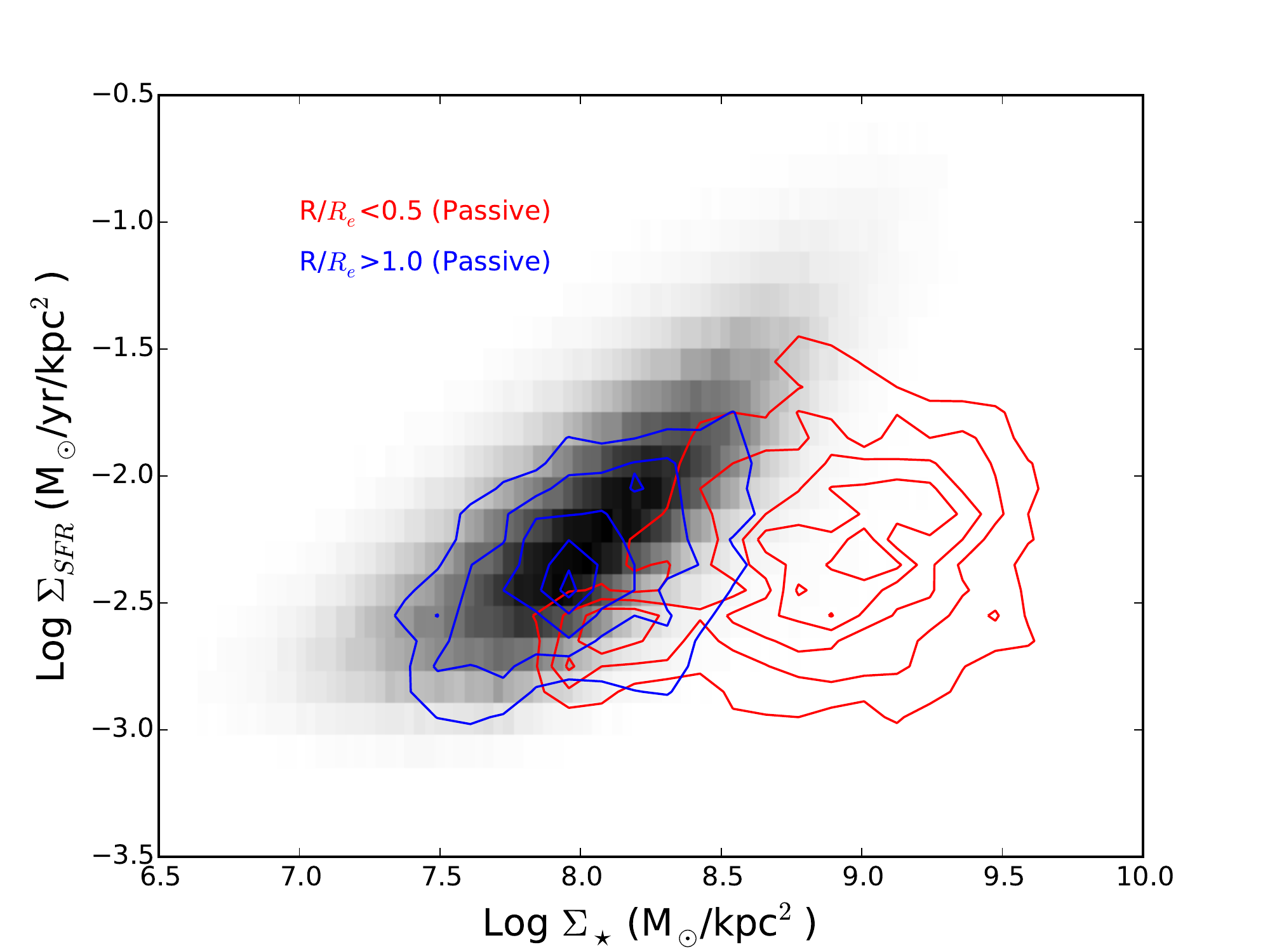}
        \caption{The grey histogram in both panels shows the resolved star forming main sequence for
          all star-forming spaxels (the sample shown in Fig. \ref{manga_sfms}).  Coloured
          contours show the distribution of \sigmass\ and \sigsfr\ for star forming spaxels in star
          forming galaxies (top panel) and star forming spaxels in passive galaxies
          (bottom) panel in two bins of $R/R_e$.  The most striking feature of this figure is the
          suppressed \sigsfr\ values in passive galaxies at $R/R_e < 0.5$, indicating that
        quenching is most dramatic in the inner galactic regions, as expected from inside-out quenching.}
    \label{ms_sf_q}
\end{figure}

We use the resolved star forming main sequence as a starting point for
our investigation of where star formation is quenched or boosted.
Hsieh et al. (2017) have recently shown that star forming spaxels
in quiescent galaxies observed with MaNGA lie below the resolved main sequence;
here we extend that work by looking at a radial dependence of that
suppression.

In Fig. \ref{ms_sf_q} we again show the distribution of all $\sim$ 487,000
star forming spaxels in the MaNGA DR13 sample.  In addition,
we show in the distributions of spaxels for just the star forming
galaxies (upper panel) and just the passive galaxies (lower panel) split
into two radial subsamples of $R/R_e<0.5$ and $R/R_e > 1.0$ in red and blue
contours respectively (where $R$ is measured from the IFU centre).  
As expected from known mass profiles (e.g.
Gonz\'{a}lez Delgado et al. 2014, 2015), both galaxy samples are dominated 
by high \sigmass\ at low
radii and low \sigmass\ at large radii.  The star forming galaxies
do not show any obvious offset from the resolved main sequence as
a function of radius.  This is perhaps not surprising as the star
forming galaxy sample
contains galaxies both above and below the main sequence (Fig. \ref{ms_dsfr}),
such that opposing trends are likely to be present.  The main result
of Fig. \ref{ms_sf_q} is conveyed in the lower panel in which it is
seen that spaxels at $R/R_e<0.5$
in passive galaxies lie far below the resolved main sequence, with
\sigsfr\ almost an order of magnitude lower than expected for their
\sigmass.  Conversely, the outer spaxels of passive galaxies appear to lie mostly
on the main sequence, with only a minority (as shown by the outer blue
contour in the lower panel of Fig. \ref{ms_sf_q}) filling the region
between the sequence and the sample sensitivity.  Fig. \ref{ms_sf_q}
therefore provides evidence that the inner regions of passive galaxies
are experiencing a preferential suppression of star formation, or `inside-out'
quenching.  However, a more detailed radial
decomposition of offsets from the resolved main sequence is needed
in order to reveal the full nature of star formation quenching and boosting.
In the next sub-section, we will quantify the metric developed for
this task.

\subsection{\dsfrs\ offsets for spaxels}\label{dsfrs_sec}

The existance of a local-scale star forming main sequence means that it is possible
to compute the offset between each spaxel's measured \sigsfr\ and that of a
set of matched control spaxels to yield a \dsfrs, in an analogous way to our 
calculation of a global \dsfr.   The pool of spaxels used to construct the
bespoke control sample for any given spaxel includes all $\sim$ 487,000 star-forming
spaxels (defined above).  

In computing the global \dsfr, we made our control sample by matching in stellar
mass, redshift and local density, under the assumption that these
parameters could modulate changes in the galactic total SFR. 
For the spaxel \dsfrs\ calculation, we must again assess the relevant
matching parameters.  By definition, an offset from the resolved
main sequence must at least be matched in $\Sigma_{\star}$, for which
we adopt a matching tolerance 0.1 dex.  To account for higher order
(i.e. not simply related to exponential \sigmass\ profiles, e.g.
Gonz\'{a}lez Delgado et al. 2014, 2015; S\'{a}nchez et al. 2017b) radial gradients we additionally 
match in the spaxel position, which we quantify
via the radial distance, $R$, from the IFU centre in units of $r$-band half 
light (effective) radius ($R_e$, taken from Simard et al. 2011).  The radial distance
of a given spaxel is matched to controls within $\pm$ 0.1 $R_e$.  We note,
however, that this radial matching does not appear to play a significant
role as we recover qualitatively similar results without radial
matching.

\begin{figure*}
	\includegraphics[width=\columnwidth]{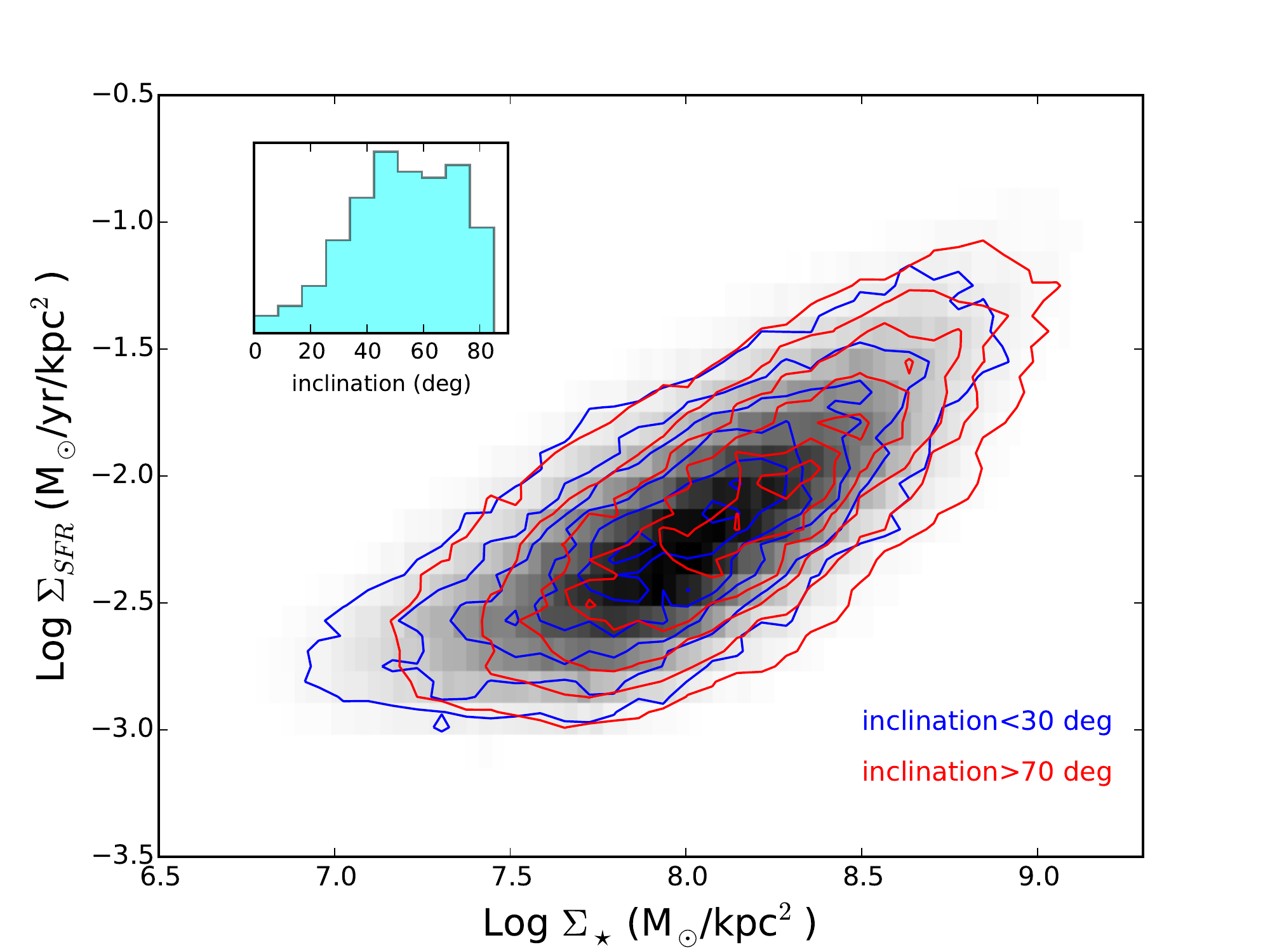}
	\includegraphics[width=\columnwidth]{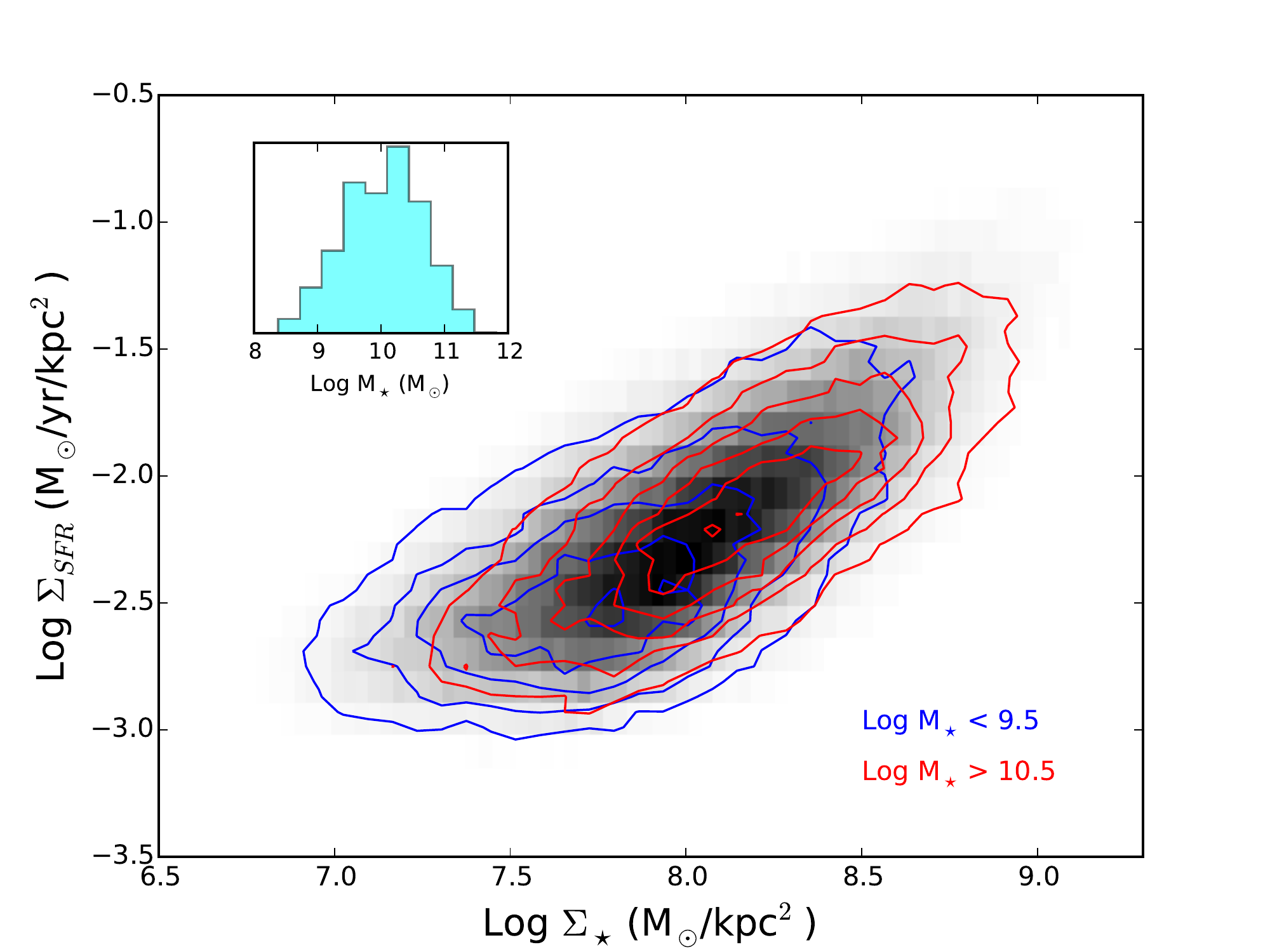}
	\includegraphics[width=\columnwidth]{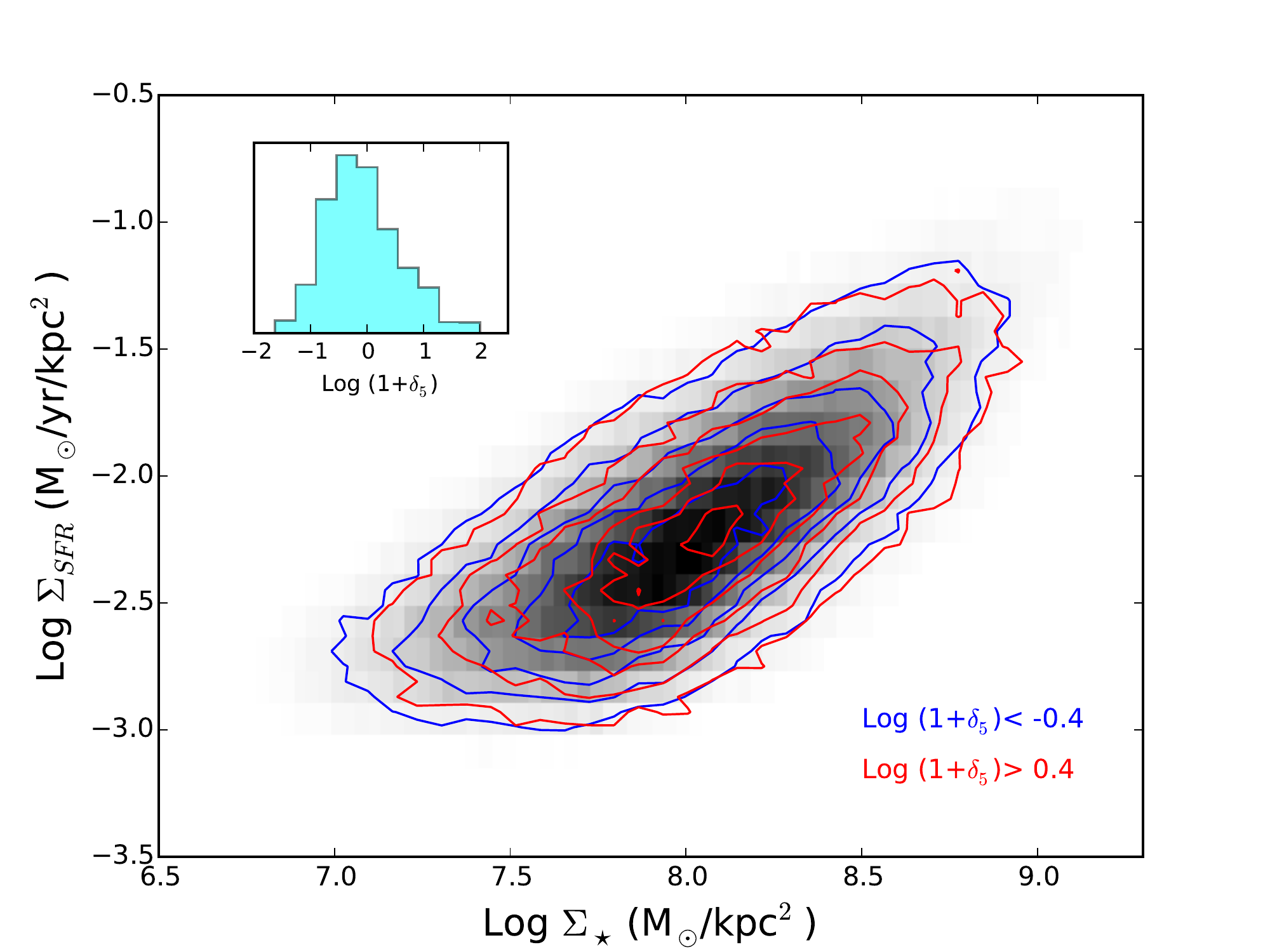}
	\includegraphics[width=\columnwidth]{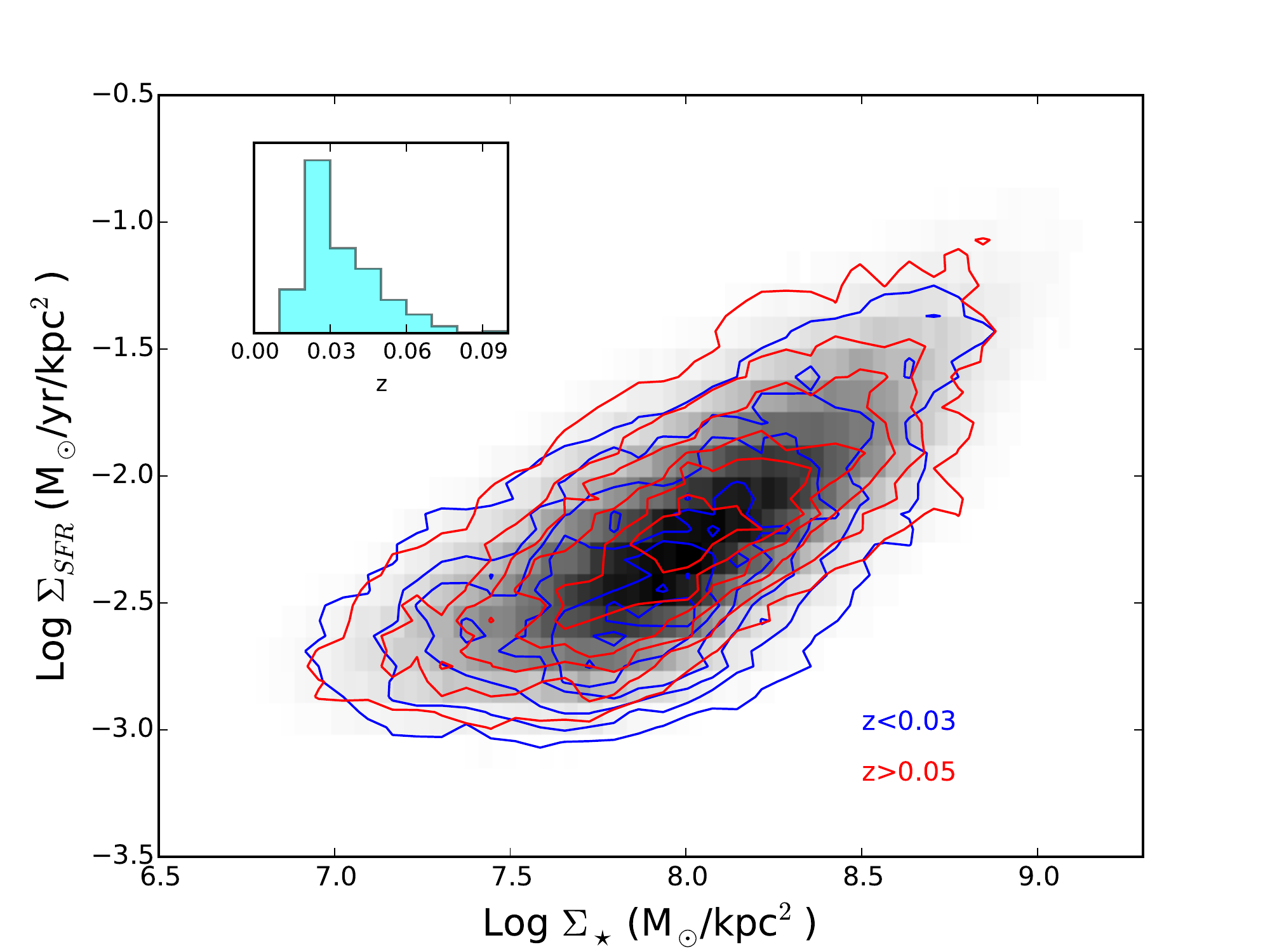}
        \caption{The local (`resolved') star forming main sequence for star-forming spaxels 
in the MaNGA DR13 datacubes.  The grey histogram (all panels) contains all star-forming 
spaxels (i.e. the same sample shown in Fig. \ref{manga_sfms}).  The blue and
red contours in the various panels show the resolved main sequence for two cuts in
inclination (top left), M$_{\star}$ (top right), log ($1 + \delta_5$) (lower left) and
$z$ (lower right).  Inset histograms show the full range of inclination, stellar
mass, $\delta_5$ and $z$ in the sample.  The slope of the resolved main sequence is independent of
inclination, M$_{\star}$, $\delta_5$, and $z$, within the ranges contained in our sample.}
    \label{ms_bt}
\end{figure*}

We investigate the need to match on global galaxy
parameters by looking for a dependence on the resolved star forming
main sequence within these parameters.  In Fig. \ref{ms_bt}
we show the main sequence
for the full sample of star-forming spaxels in the grey 2-d histogram and
in coloured contours the distribution for two bins of inclination (top left panel),
total stellar mass (top right panel), environment (bottom left panel) and redshift
(bottom right panel).  
No dependence of the resolved main sequence is found for any
of the tested properties, leading us to
conclude that, for the range of properties in our sample, the
resolved main sequence is invariant to changes in these properties.

One galaxy property that \textit{has} been previously shown to
impact the resolved main sequence is morphology, wherein galaxies
with higher bulge fractions tend to exhibit lower \sigsfr\ for
their \sigmass\ (e.g. Gonz\'{a}lez Delgado et al. 2016;
Maragkoudakis et al. 2016).  However, we do not match in galaxy
B/T in our \dsfrs\ calculation for several reasons. First, our
sample is dominated by galaxies with B/T $<$ 0.4, a regime
in which morphology does not strongly affect the resolved
main sequence.  Second, Pan et al. (in prep)
have shown that truly star forming spaxels, selected using emission
line diagnostics in a similar way to our procedure described
above, show relatively little dependence on bulge fraction.
Instead, Pan et al. (in prep) conclude that resolved main
sequence offsets for high B/T galaxies are dominated by
spaxels ionized by other processes.
Finally, the global main sequence shows a similar dependence on
structure, in which bulge dominated galaxies
tend to have low SFR for their M$_{\star}$ and have a higher
quenched fraction (e.g. Wuyts et al. 2011b; Bluck et al. 2014).
In this sense, the global main sequence is once again an
extension of the kpc-scale relationships.  We therefore do
not match spaxels based on their parent galaxy's morphology,
since morphology itself appears to correlate with \dsfr,
but return to investigate the dependence on B/T explicitly in Section
\ref{profiles_sec}.

The assembly of the control spaxels thus entails matching on \sigmass\
and distance from the galaxy centre (in units of $R_e$) and is computed as:

\begin{equation}
  \Delta \Sigma_{SFR} = \log \Sigma_{\rm SFR, spaxel} - \log \Sigma_{\rm SFR, control}.
\end{equation}

As for the global \dsfr, $\Sigma_{\rm SFR, control}$ is taken as the median
value of all of the matched control spaxels. As for the calculation of
\dsfr, we again require at least 5 spaxels to be matched in order to consider
the control matching successful. If fewer than five spaxels are matched the
tolerances are iteratively grown by a further 0.1 dex
and 0.1 in $\Sigma_{\star}$ and $R/R_e$ respectively.  However, in practice the very large
control pool of star-forming spaxels means that $>$99 percent of
spaxels have the required limit of 5 matched spaxels
without the need to grow the matching tolerances.  The mean number of
matched control spaxels to any given spaxel is $\sim$ 6000.

In closing this section, we note that \dsfrs\ theoretically captures the same relative
difference in star formation as profiles of spaxel specific SFR (sSFR, e.g.
Gonz\'{a}lez Delgado et al. 2015; Belfiore et al. 2017b; Spindler et al. 2017;
Morselli et al. in prep), since both measure a SFR
relative to a mass.  However, there are two reasons we adopt \dsfrs\
instead of sSFR in this work.  First, with the \dsfrs\ metric
we are able to additionally
control for any extra parameters of interest; in our definition
of \dsfrs\ we control for radius as well as mass surface density.  Second,
a differential analysis, which computes a \sigsfr\ relative to
a matched control, mitigates spaxel selection biases.  Due
to the S/N criterion of our star forming spaxel sample, we
are incomplete for low \sigsfr\ at low \sigmass.  Consequently,
the mean spaxel sSFR is biased to high values at low \sigmass\
which preferentially occur at large radii, and subsequently
alter the radial profiles.  Further issues related to spaxel
selection biases are discussed in Section \ref{bias_sec}.

\subsection{$\Delta$ O/H offsets for spaxels}\label{doh_sec}

\begin{figure}
	\includegraphics[width=\columnwidth]{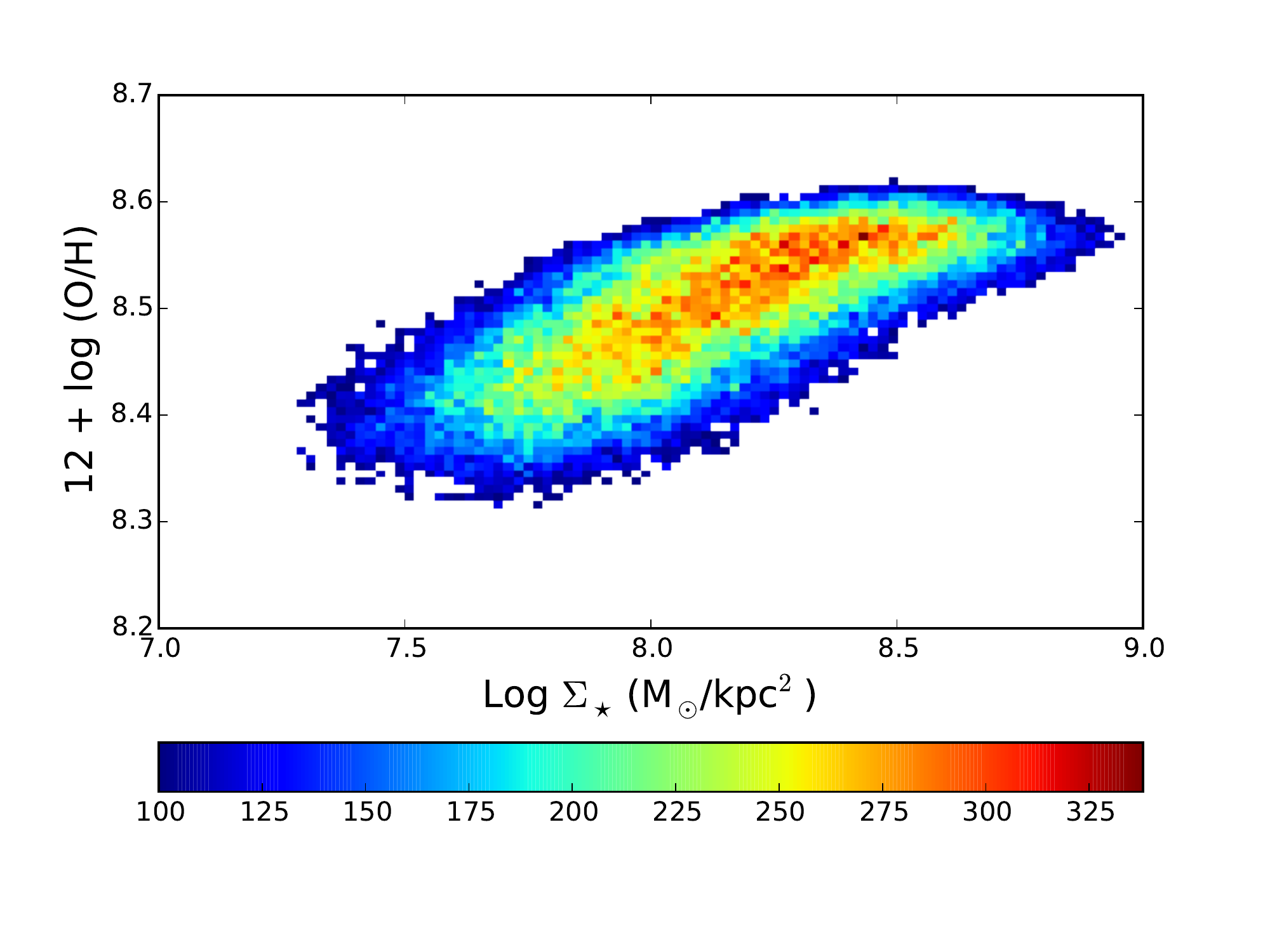}
        \caption{The local (`resolved') mass metallicity relation for $\sim$487,000
          star-forming spaxels in the MaNGA DR13 datacubes.
          The colour bar indicates the number of spaxels in 
each bin.   }
    \label{manga_mzr}
\end{figure}

Just as the star forming main sequence is a global rendering of
the resolved relation (Fig \ref{manga_sfms}), the global mass
metallicity relation (MZR, e.g. Tremonti et al. 2004; Ellison et al. 
2008b) is recovered on local scales (Moran et al. 2012; Rosales-Ortega
et al. 2012; Barrera-Ballesteros et al. 2016).  
Following Barrera-Ballesteros et al. (2016),
the gas phase metallicities are computed for each spaxel using
the calibration of Marino et al. (2013), which is based on a
large compilation of direct electron temperature ($T_e$) abundance measurements.
In particular, the Marino et al. (2013) calibration extends previous
efforts to calibrate $T_e$ abundances against strong emission lines
(e.g. Pettini \& Pagel 2004) by including \HII\ regions that
extend to higher metallicity.  Based on a fit to 603 \HII\ region
$T_e$ abundances and their ratios of [\OIII], [\NII], H$\alpha$ and
H$\beta$ lines, Marino et al. (2013) find a best fitting relation:

\begin{equation}\label{eqn_oh}
12 + \log (O/H) = 8.533[\pm0.012] - 0.214[\pm0.012] \times O3N2 
\end{equation}

where

\begin{equation}\label{eqn_o3n2}
O3N2 =  \log \Big( \frac{[OIII]\lambda 5007}{H\beta} \times \frac{H\alpha}{[NII]\lambda 6583} \Big).
\end{equation}

Metallicities are computed for all of the star forming spaxels in our
sample using equations \ref{eqn_oh} and \ref{eqn_o3n2} using the extinction
corrected fluxes, as described in Sec. \ref{pipe3d_sec}.
The resulting resolved mass metallicity relation for the star forming
spaxels in our MaNGA sample is shown in Fig. \ref{manga_mzr}
(as previously found by Barrera-Ballesteros et al. 2016).

Having established the local MZR for the MaNGA spaxels, we can now
compute a spaxel metallicity offset in an analogous way to
the calculation of the spaxel \dsfrs, matching each spaxel
to a control in a narrow tolerance of \sigmass\ and $R/R_e$.
Thus, the metallicity offset is defined as:

\begin{equation}
  \Delta O/H = \log \Sigma_{\rm O/H, spaxel} - \log \Sigma_{\rm O/H, control}.
\end{equation}

Once again, this differential approach mitigates biases and selection
effects.  In the case of metallicity calibrations, it is well known
(e.g. Kewley \& Ellison 2008) that different strong line diagnostics
can yield abundances that differ by almost an order of magnitude.
However, \textit{relative} abundances, within a given calibration,
are quite robust.

\section{Star formation profiles as a function of main sequence offset}

\begin{figure*}
	\includegraphics[width=14cm]{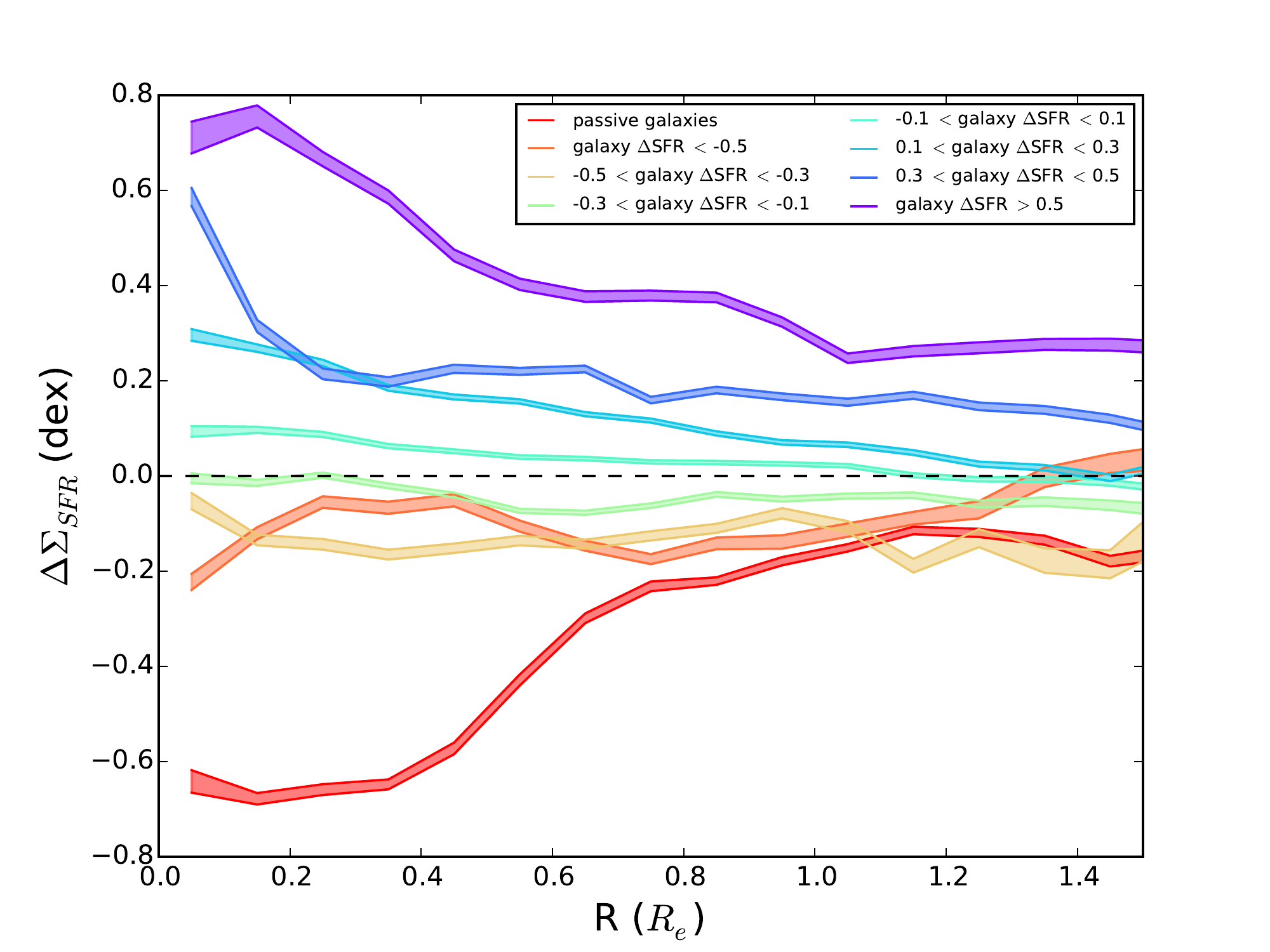}
	\includegraphics[width=14cm]{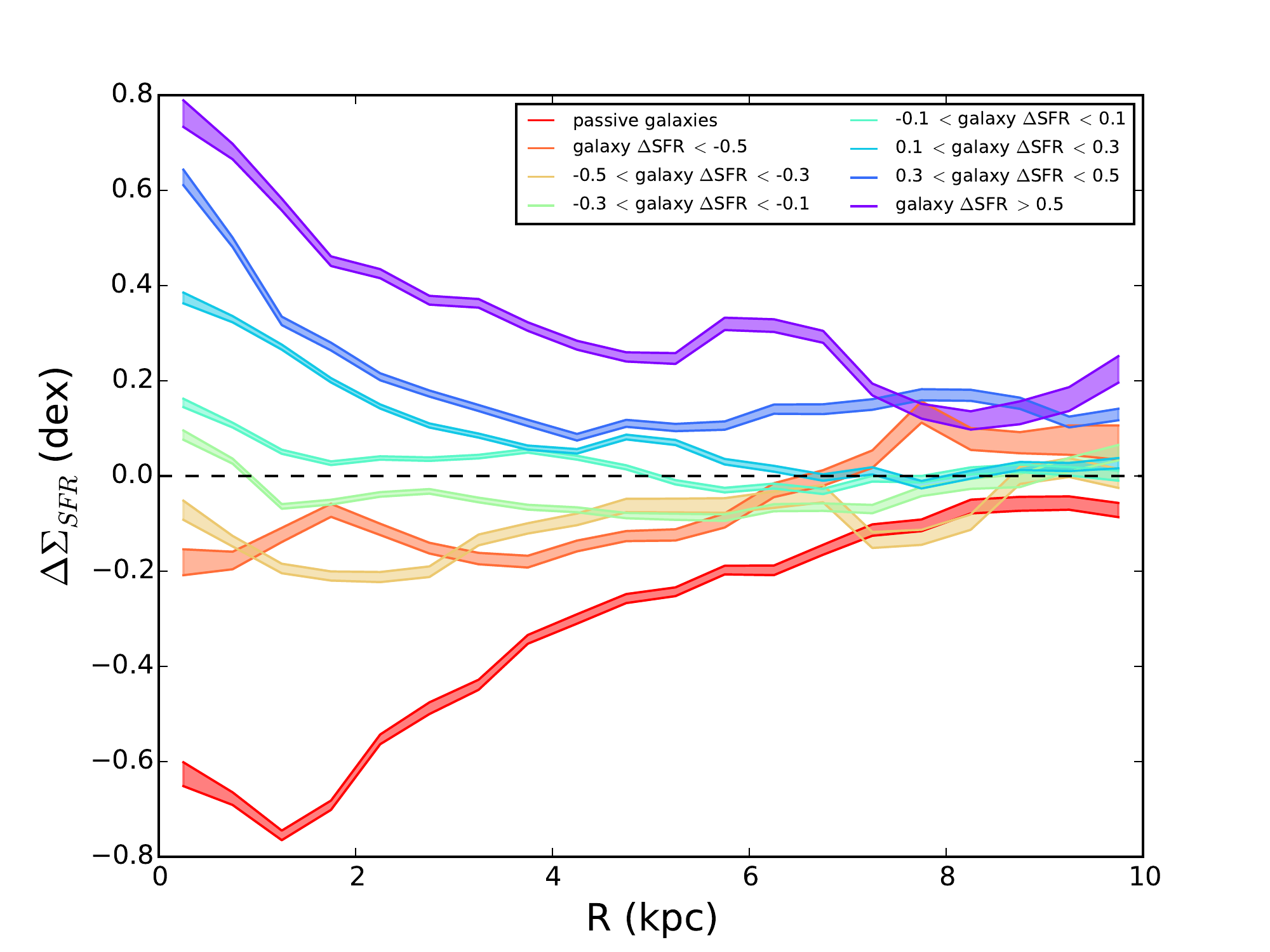}
        \caption{Radial profiles of \dsfrs\ for spaxels that inhabit galaxies with varying positions on the global main sequence (i.e. varying \dsfr).  The horizontal dashed line indicates zero enhancement or suppression of \sigsfr\ relative to control spaxels of the same $\Sigma_{\star}$ and radial distance from the galaxy centre.  The top and bottom panels show profiles on in units of $R_e$ and kpc respectively.}
    \label{ms_dsfr_profiles}
\end{figure*}

The main goal of the present work is to investigate the radial profiles
of star formation as a function of the global galaxy main sequence offset (\dsfr),
in order to gain insight into the mechanisms that modulate star formation.
Since galaxies fundamentally show radial gradients in star formation
(e.g. Gonz\'{a}lez Delgado et al. 2016), we can't use the \sigsfr\ profiles 
alone to assess where, and by how much, star formation is particularly boosted
or suppressed.  On the other hand, since
\dsfrs\ controls (for each spaxel) for both \sigmass\ (hence the
expected \sigsfr) and
radial position in the galaxy, it removes the underlying star
formation rate gradient.  Therefore, \dsfrs\ is a measure of
how much extra (or less) star formation is in a given spaxel
compared to the `norm' for its \sigmass\ and radial distance
from the centre.  

\subsection{Radial \dsfrs\ profiles}\label{profiles_sec}

In Fig.  \ref{ms_dsfr_profiles} we present median profiles of 
\dsfrs\ in bins of \dsfr\ in units of both $R_e$ (top
panel) and kpc (bottom panel).  For reference, the average half
light radius in the sample is $\sim$ 6 kpc.  The median PSF
of MaNGA observations (2.5 arcsec) corresponds to 1.5 kpc at the
median redshift of the sample ($z=0.03$), such that the radial
profiles are well resolved.
Fig. \ref{ms_dsfr_profiles} shows that galaxies that lie above the global
star-forming main sequence (positive \dsfr) exhibit elevated \dsfrs\
out to at least 1.5 times the galactic half light radius ($\sim$ 10 kpc), with the
average enhancement proportional to the global \dsfr.  Moreover, the
\dsfrs\ values increase towards smaller radii, indicating that galaxies
above the main sequence are particularly prodigious in their star formation
within the inner 0.5 $R_e$ ($\sim$ 3 kpc).   The radial
profiles of \dsfrs\ shown in Fig. \ref{ms_dsfr_profiles} indicate that
boosts in star formation are apparently regulated from the inside out.

Turning now to galaxies located below the main sequence.  In contrast
with the significant (several tenths of a dex) galaxy-wide
enhancement of star formation in positive \dsfr\ galaxies, star
forming galaxies below the main sequence exhibit modest suppression
of \dsfrs.  Even in our most extreme bin of \dsfr\ $<-0.5$, the
radial average profile of \dsfrs\ does not drop below \dsfrs\ $\sim -0.2$ dex.
There is also no strong radial dependence of \dsfrs\ in star-forming
galaxies that lie below the main sequence.  Our results are therefore 
consistent with Belfiore et al. (2017b) whose sample of `green valley' galaxies,
which show fairly uniformly suppressed sSFRs and are qualitatively similarly to 
our `below main sequence star forming' galaxies. However, the passive
galaxy population (whose equivalent \dsfr\ $<-1.0$) does show
a strong radial \dsfrs\ profile.  At large radii (beyond the half
light radius) the suppression of star formation is mild, only 0.2 dex
and consistent with star forming galaxies that are only a factor of a 
few below the main sequence.  However within $\sim$ 0.5 $R_e$ ($\sim$ 3
kpc) the star formation is suppressed by a factor of 4, a suppression
which mirrors the enhancement seen in the galaxies located in
the highest slice above the main sequence.  The strong central
SFR suppressions in the passive population are qualitatively
similar to the `centrally suppressed' galaxies studied by Spindler
et al. (2017).

\begin{figure}
	\includegraphics[width=\columnwidth]{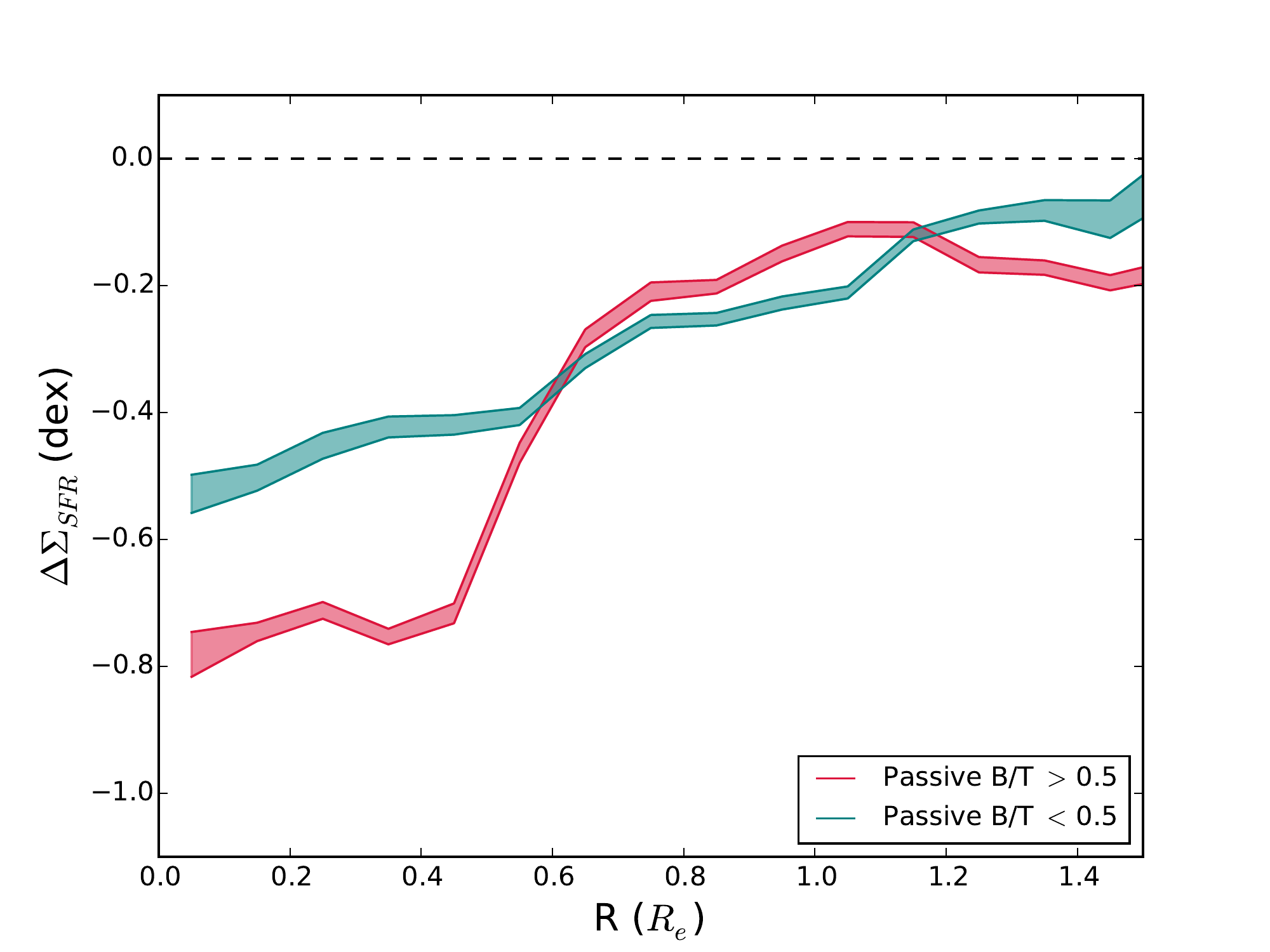}
        \caption{Radial profiles of \dsfrs\ for spaxels in disk dominated (teal line) and bulge dominated (crimson line) passive galaxies.  Bulge fractions are determined from the bulge and disk mass catalog of Mendel et al. (2014).  The horizontal dashed line indicates zero enhancement or suppression of \sigsfr\ relative to control spaxels of the same $\Sigma_{\star}$ and radial distance from the galaxy centre.  }
    \label{dsfr_bt}
\end{figure}

The trends in \dsfrs\ profiles for \textit{star-forming galaxies} in Fig.  \ref{ms_dsfr_profiles} 
are not driven by variations in bulge fraction.  Gonz\'{a}lez Delgado
et al. (2016) have shown that the \sigsfr\ profiles of
disk dominated galaxies show little dependence on their detailed morohology (bulge fraction). 
For the star-forming sample studied here, 85 per cent of galaxies have
$r$-band B/T$<$0.4.  However, the passive sample has a broader
range of B/T; approximately 2/3 of the 470 galaxies in that sample
have B/T $>$ 0.5.  In Fig. \ref{dsfr_bt} we separate the passive population
into bulge dominated (B/T$>$0.5) and disk dominated (B/T$<$0.5) sub-samples.
As discussed in Tachella et al. (2015b), radial dependences
on bulge fraction can be potentially misleading when B/T is measured in optical light.
We have therefore defined our bulge fractions in mass, using the bulge and
disk mass catalog of Mendel et al. (2014).
From Fig.  \ref{dsfr_bt} it can be seen that the passive galaxies in both
morphological bins show similar \dsfrs\ profiles beyond $\sim$ 0.6 $R_e$.
However, passive galaxies that have assembled a significant bulge have
a factor of two lower central \dsfrs, compared with disk dominated passive galaxies.
The \dsfrs\ profiles of passive galaxies therefore appear to be dependent
on the presence of a bulge.  

\subsection{Spaxel selection biases}\label{bias_sec}

We have checked that the \dsfrs\ profile for passive galaxies is not an
artefact of the spaxel selection process, in which we have imposed
a S/N limit of 3, which in turn limits the \sigsfr\ threshold of
the sample (Fig \ref{manga_sfms}).  A particular concern may be that,
at low \sigmass\ (typically found in the outer disk) our selection preferentially
excludes low \sigsfr\ spaxels and could hence bias the profile to
large values at large radii.  This effect should be largely mitigated
by our differential approach of a matched comparison sample, in which
control spaxels are subject to the same bias.  Nonetheless,
we have repeated the \dsfrs\ profile analysis
with a less aggressive S/N threshold, requiring only that the
spaxel be below the Kewley et al. (2001) demarcation, H$\alpha$
EW exceeds 6\AA\ and that the H$\alpha$ and H$\beta$ S/N $>$ 1.  
The combination of the Kewley et al. (2001) AGN
criterion with an H$\alpha$ EW cut has previously been used
as a more inclusive selection for star forming spaxels
(e.g.  S\'{a}nchez et al. 2014;
P\'{e}rez-Montero et al. 2016; S\'{a}nchez-Menguiano et al. 2016).
Relaxing the S/N threshold effectively extends the \sigsfr\
threshold of the resolved main sequence by almost 1 dex, such that the
lowest star formation rate surface densities extend to \sigsfr\ $\sim -4$.  
Despite this effective increase in sensitivity to low \sigsfr\ the
profiles of \dsfrs\ remain qualitatively unchanged.  Importantly, 
the \dsfrs\ values at large radii are robust.  The main difference in
the \dsfrs\ profiles when we adopt an H$\alpha$ EW limit is that the
suppression of star formation in passive galaxies at small radii is
reduced from $\sim -0.7$ dex to $\sim -0.4$ dex.  This is because the
H$\alpha$ EW is equivalent to a specific SFR, such that a cut at H$\alpha$
EW = 6 \AA\ is approximately a cut of log sSFR $\sim -11$ yr$^{-1}$.  Therefore,
an H$\alpha$ EW cut introduces a bias against spaxels with low sSFR
surface density and spaxels signficantly below the main sequence 
(such as those at small radii in passive galaxies, as shown by the red
contours in Fig  \ref{ms_sf_q}).  For the purposes of our analysis,
a spaxel selection based on S/N cut is therefore more sensitive to
the central suppresion of star formatin than a combination of the
Kewley et al. (2001) threshold and an H$\alpha$ EW cut.

As another test of possible selection bias, we have also repeated 
the profile analysis
including only spaxels above log \sigmass\ = 8 M$_{\odot}$/kpc$^2$, above
which the \sigsfr\ selection on the resolved main sequence
should be fairly complete (see Fig. \ref{manga_sfms}).
Again, the \dsfrs\ profile shape
for the passive galaxies is qualitatively similar.  Finally, we
refer the reader back to Fig. \ref{ms_sf_q} in which the spaxels
at large $R$ in passive galaxies (blue contours in the lower
panel) are mostly on the main sequence, even though the parameter
space is sensitive to lower values of \sigsfr.  We conclude that the
positive gradient in \dsfrs\ for passive galaxies in Fig.
\ref{ms_dsfr_profiles} is not a result of selection biases, but
reflects a true relative decrease in the star formation towards
the centres of galaxies that lie well below the main sequence.

\section{Discussion}

The main goal of the current work has been to quantify the radial
profile of star formation rate regulation at $z \sim 0$.
We have made a careful definition of a galaxy's global 
star formation rate relative to the `norm', by computing a
\dsfr\ relative to stellar mass, redshift and local environment average.
We also introduce a new metric, \dsfrs, which measures a spaxel-based 
star formation enhancement or deficit relative to the resolved main
sequence, permitting a radial assessment of where star formation
is boosted or quenched.  We now review our results in the context
of the mechanisms that regulate the radial star formation profiles
and other work in the literature.

\subsection{Positive offsets from the main sequence - where do star bursts happen?}

In terms of galaxies that lie above the global main sequence, the primary result
of this paper is that elevated \sigsfr\ is present throughout the galaxy,
with the greatest enhancements in the central regions (purple and blue
profiles in Fig \ref{ms_dsfr_profiles}).  Our result is consistent with
the SDSS study of Morselli et al.
(2017) who use bulge and disk photometry of SDSS galaxies to conclude that
galaxies above the main sequence require both star forming
disks \textit{and} star forming bulges. 

\begin{figure*}
	\includegraphics[width=15cm]{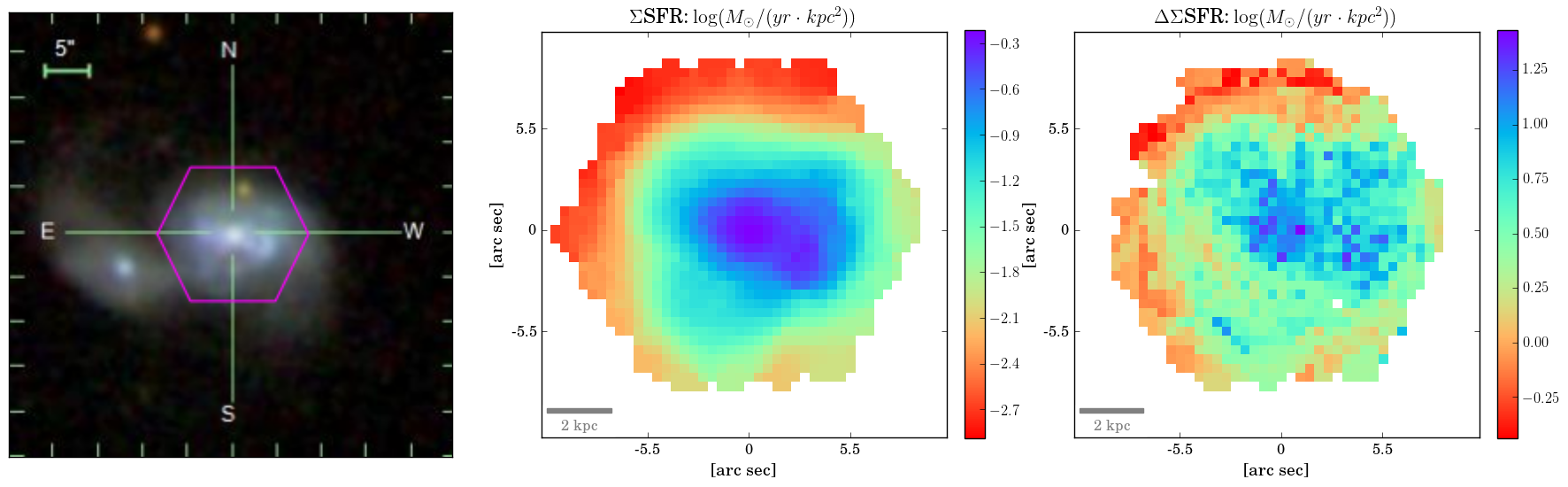} 
	\includegraphics[width=15cm]{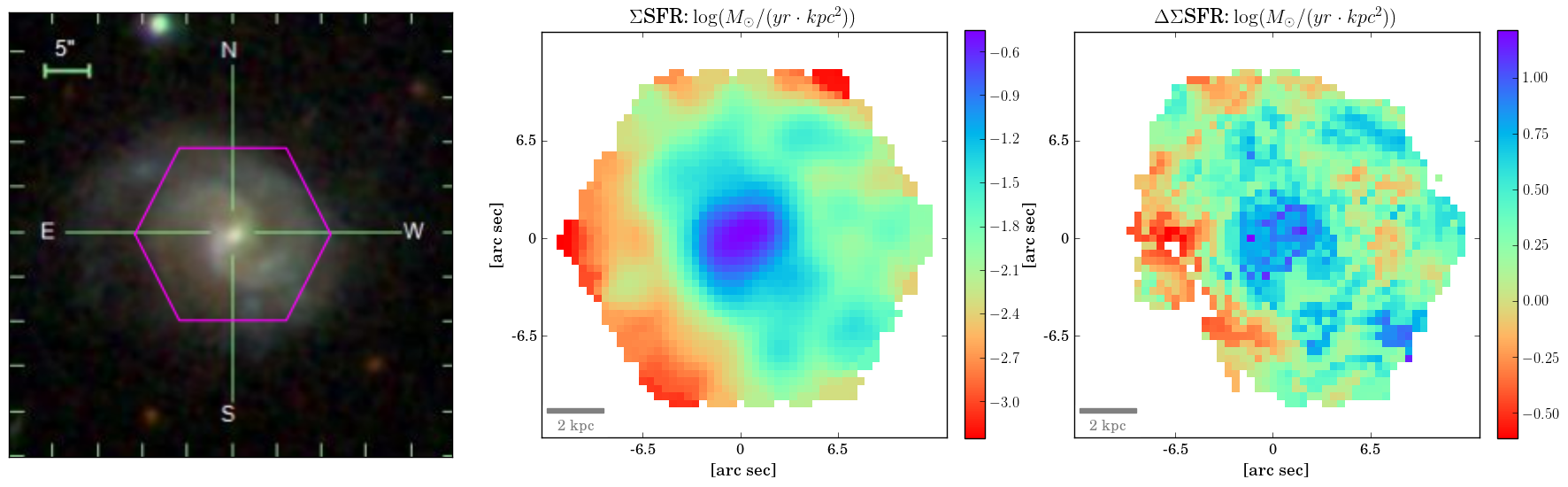} 
        \caption{Examples of mergers within the MaNGA sample.  Panels, from left to right, show
the SDSS \textit{gri} image with the MaNGA footprint overlaid as a magenta hexagon, map
of \sigsfr\ and \dsfrs.  Colour bars are not identical for each panel.  Top row: The MaNGA target
has \dsfr\ = 1.11 dex and a bright companion just outside the IFU footprint.  Bottom row:
The MaNGA target has \dsfr\ =0.39 dex and a faint companion outside the IFU footprint,
towards the north east.  Both examples show clear enhancement in their central star
formation (right panels).  }
    \label{merger_map}
\end{figure*}

Several recent studies at moderately high redshifts have similarly
concluded that elevated star formation is widespread in galaxies that
lie above the main sequence.  Magdis et al. (2016)
have shown that \sigsfr\ correlates with main sequence
offset at $z \sim 1$, although they do not investigate the radial
dependence of the elevation.  Nelson et al. (2016) further showed
that the elevation in star formation for galaxies lying above the
main sequence at these redshifts was roughly constant (a factor of $\sim$ 2)
throughout the disk on scales of 2--6 kpc (see also Morselli et al.,
in prep), and is reproduced by simulations of galaxies with burtsy
histories (Orr et al. 2017).  These results are consistent with our 
result in Fig \ref{ms_dsfr_profiles} that show elevated \dsfrs\ out to 
at least 10 kpc.  However, we additionally find that the
profile of star formation enhancement increases further in the central 3 kpc
(0.5 $R_e$).  Although the Nelson et al. (2016) sample is at considerably
higher redshift than ours, another possible reason for the apparent
discrepancy between the relative SFR enhancements in the central region
could be the role of dust, which is unaccounted for in the Nelson
et al. (2016) study.  Galaxies
above the main sequence are characterized by elevated levels of
both star formation and dust (Wuyts et al. 2011b; Whitaker et al. 2012).
Hence, the uncorrected H$\alpha$ fluxes
may under-estimate the total SFR (Wuyts et al. 2011a,b), which may be
a particular issue in the central regions of highly star forming galaxies. 
Finally, a high redshift ($z \sim$ 2) analog of our results is presented
by Tacchella et al. (2017) who find that galaxies above the main sequence
have higher sSFRs in the inner 3 kpc than at larger radii, consistent with
expectations of gas inflow models (Tacchella et al. 2016a,b).

\subsubsection{The role of mergers}

The observation of enhanced \dsfrs\ in the central regions of galaxies is consistent
with theoretical expectations of triggered star formation in galaxy mergers
 (e.g. Barnes \& Hernquist 1991;  Mihos \& Hernquist 1994, 1996; Torrey et al. 2012; 
Moreno et al. 2015).  Indeed, observations with both
single fibre (e.g. Ellison et al. 2013) and IFU data (Barrera-Ballesteros et al.
2015; Cortijo-Ferrero et al. 2017a,b)
have confirmed that SFR enhancements in galaxy mergers can be widespread, but
are statistically centrally located.  This is further supported by
CO observations of merging galaxies that find compact central molecular gas 
disks from which the starburst is fed (e.g. Ueda et al. 2014;
Yamashita et al. 2017).  A visual inspection of the galaxies with the largest \dsfr\ enhancements above
the global main sequence reveals that some are clearly galaxy mergers.
In Fig. \ref{merger_map} we show two such examples.  Panels in the upper row
show the SDSS image, map of \sigsfr\ and map of \dsfrs\ (from left to right)
of a galaxy with a close companion (outside of the MaNGA footprint, shown by the
magenta hexagon), whose \dsfr\
is +1.11 dex, one of the highest main sequence offsets in our sample.  The
lower panels in Fig \ref{merger_map} represent the same quantities for another
galaxy merger with \dsfr\ = +0.39 dex.  Evidence for an interaction with a much fainter
companion can be seen outside of the MaNGA footprint towards the north east.
The \dsfrs\ maps (right panels) for both galaxies clearly show that the central regions 
exhibit the greatest star formation rate enhancements.
A study focusing specifically on the star formation rate profiles of galaxies 
in mergers will be presented in a forthcoming work.

Despite the presence of mergers in the MaNGA sample, overall they are
in the minority.  We performed a visual classification of galaxies
that either have a close companion or show obvious signs of interaction (including
postmergers: single galaxies with signs of disturbance).  Of the 392
star forming MaNGA galaxies used in this work, only $\sim$ 50 have an obvious
companion or are classified as a post-merger
We have repeated our analysis of \dsfrs\ profiles excluding galaxies that have either
been identified as a possible merger by our visual classification, or with a strict
merger vote fraction cut (pMerger$<$0.05) based on Galaxy Zoo (Lintott et al. 2008;
Darg et al. 2010).  There is no significant change in our results when spaxels located
in galaxies engaged in an interaction are excluded, indicating that (in general)
other processes
drive the centrally enhanced \sigsfr\ profiles in Fig. \ref{ms_dsfr_profiles}. 
Nonetheless, it remains possible that the accretion of external gas
still plays an important role in triggering centrally concentrated star
formation, since minor mergers, interactions with dwarf satellites and
smooth accretion would be difficult to identify visually (and are likely
much more frequent than major mergers).  For example, Chen et al. (2016)
have inferred the accretion of external gas in nine blue galaxies in the
MaNGA sample from their counter-rotating gas kinematics.  These galaxies
are characterized by high central SFRs, but without any obvious sign
of an on-going or recent merger.

\subsubsection{In the context of the compaction model}

Simulations of high redshift galaxies have recently been used to conclude that a variety of
processes can lead to intense gas inflow events leading to compact central
star formation (e.g. Dekel \& Burkert 2014; Zolotov et al. 2015; Tacchella et al. 2016 a, b),
including mergers of varying mass ratios, streams and tidal compression.
Although the process of compaction is expected to operate most
dramatically at high redshifts, where both the merger rate and disk
gas fractions are higher than the present day,
our $z \sim 0$ results qualitatively match the expected centrally
peaked star formation in galaxies above the main sequence (e.g. Fig. 11 of Tacchella
et al. 2016a).  However, the compaction model predicts the centrally
enhanced star formation to be accompanied by a reduction (or at most, consistent)
star formation in the extended disk (e.g. Tacchella et al. 2016b).
This is not seen in our observations: galaxies that lie above the global
main sequence have elevated \sigsfr\ (i.e. positive \dsfrs)
throughout the disk.  We note that
since we are radially averaging the profiles (e.g. in Fig. \ref{ms_dsfr_profiles})
the star formation enhancements in the disk regime are not necessarily uniform.
Indeed, the positive \dsfrs\ beyond $R_e$ are often due to 
clumps of enhanced star formation at a few specific sites within the disk,
similar to clumps in higher redshift galaxies (e.g. Wisnioski et al.
2011; Wuyts et al. 2012, 2013).  Averaged together radially, these localized
sites of enhanced star formation manifest as an elevated platform
of star formation, similar to that seen in $z \sim 1$ galaxies
(Nelson et al. 2016).

The inflow of gas that precedes the central starburst in a
`wet compaction' event might be expected leave an imprint on the gas phase
metallicity of the galactic interstellar medium (ISM).  Most of the
mechanisms that trigger the gas inflow are expected to lower the
central metallicity.  For example, mergers and disk instabilities
channel gas from the outer, more metal-poor disk towards the centre
(Rupke, Kewley \& Barnes 2010; Perez, Michel-Dansac \& Tissera 2011; 
Sillero et al. 2017).
Accretion of satellites (or minor mergers) and intergalactic streams
are also expected to deliver relatively metal-poor gas (e.g. Finlator
\& Dav\'e 2008).  Such inflow models may explain the dependence of
the mass metallicity relation on SFR (e.g. Ellison et al. 2008b)
and the dependence of galaxy metallicity gradients on specific SFR
(Stott et al. 2014).

\begin{figure}
	\includegraphics[width=\columnwidth]{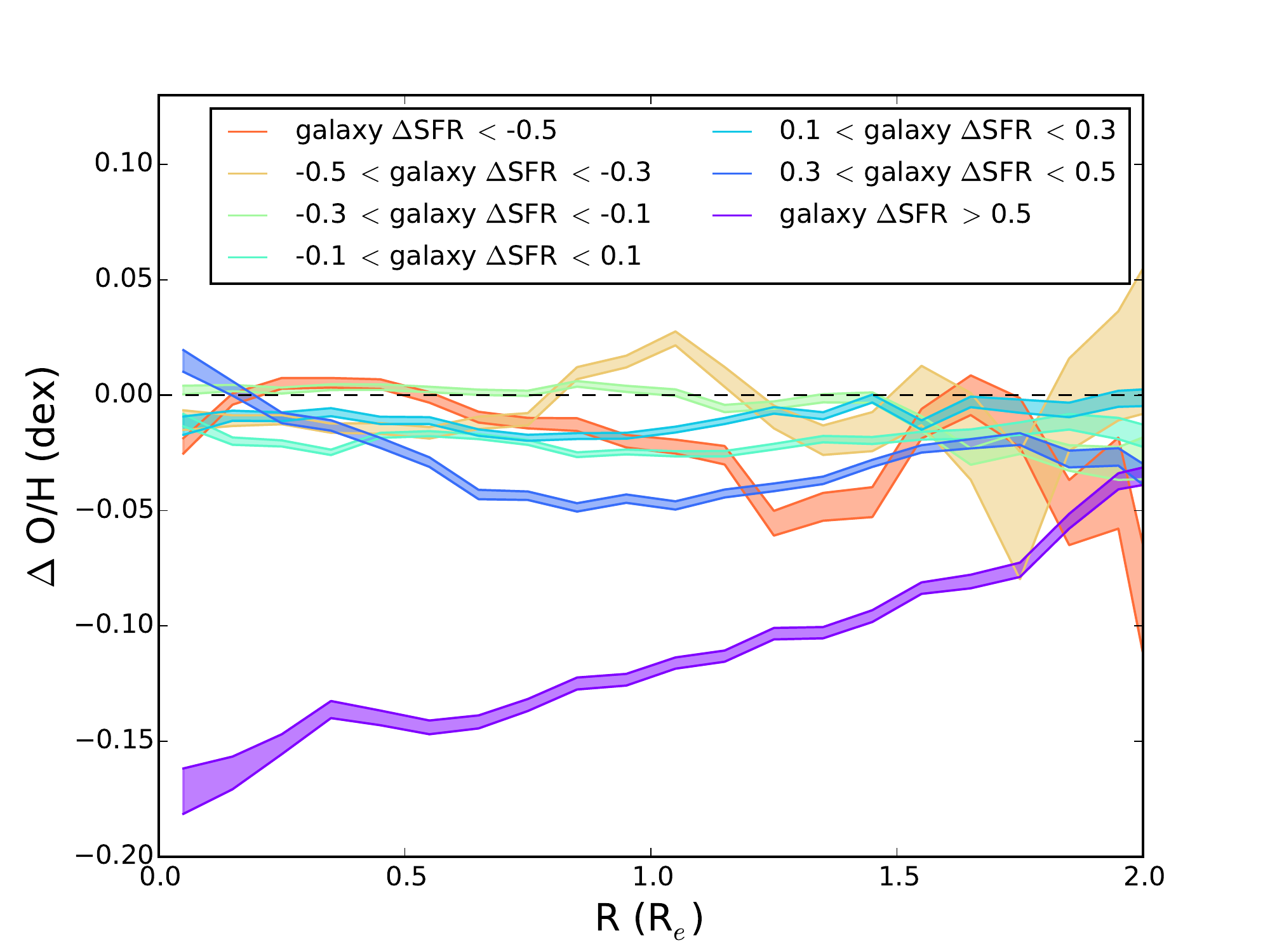}
	\includegraphics[width=\columnwidth]{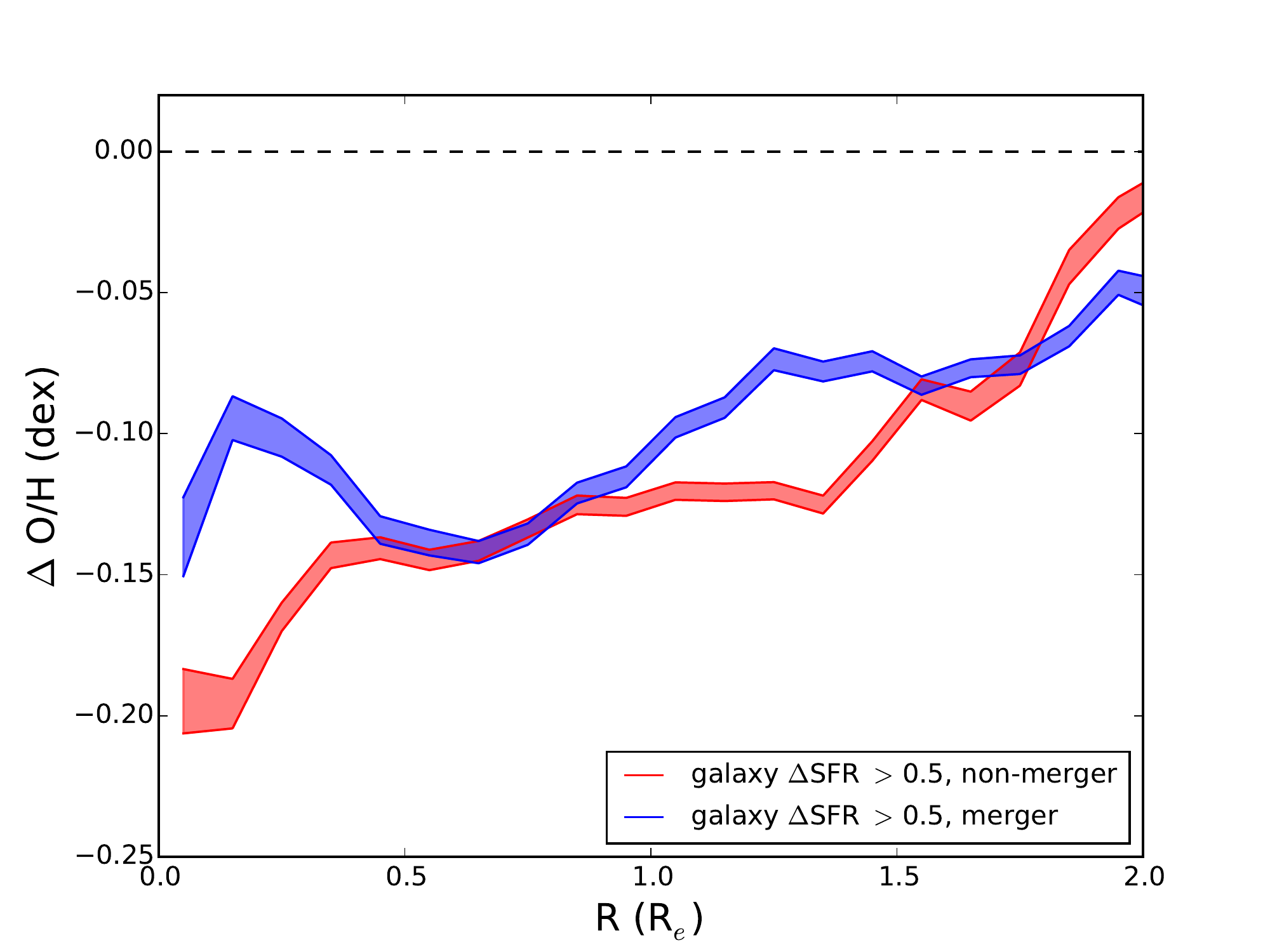}
        \caption{Top panel: Radial profiles of $\Delta$ O/H for spaxels that inhabit galaxies with varying positions on the global main sequence (i.e. varying \dsfr).  Bottom panel:  Only galaxies in the highest \dsfr\ bin (purple line in the top panel) separated into galaxies in likely mergers (either a close companion or obvious tidal disturbance) and non-mergers.  The horizontal dashed line indicates zero enhancement or suppression of O/H relative to control spaxels of the same $\Sigma_{\star}$ and radial distance from the galaxy centre. }
    \label{doh}
\end{figure}

For the star-forming galaxies in our sample, in the top panel of
Fig. \ref{doh} we plot the radial distribition of $\Delta$ O/H in
bins of \dsfr.  This plot is analogous to the profiles of \dsfrs\ in
Fig. \ref{ms_dsfr_profiles}.
The expected central dilution of metallicity is clearly present in
the galaxies that are furthest offset above the main sequence (by
at least a factor of three, purple line in the top panel Fig. \ref{doh}).  The
metallicity dilution shows a steady radial profile, increasing
from $\sim -0.2$ dex in the central region to an almost `normal'
metallicity at 2 $R_e$.  This widespread dilution indicates that
if these galaxies are diluted as a result of metal-poor gas inflow, the
source of that gas must either be external, or originating
from beyond 2 $R_e$.  Of the 21 galaxies in this highest \dsfr\
bin, 6 have either a close companion (including the example
in the top row of Fig. \ref{merger_map}) or show tidal features indicative
of a recent merger, a process previously reported to lead to central metallicity
dilution (e.g. Kewley et al. 2010; Rich et al. 2012; Cortijo-Ferrero et al.
2017a).  In order to test whether the positive metallicity gradients
seen in the highest \dsfr\ bin of Fig. \ref{doh} is driven by merging
galaxies, in the lower panel of that figure we
distinguish the mergers and non-mergers in the \dsfr\ $>0.5$
bin.   Both the mergers and non-mergers
show similar dilution profiles, indicating that a process other
than a recent (major) merger can be responsible for both the starburst
and metallicity dilution of these high \dsfr\ galaxies.  Based on
their simulations, Sillero et al. (2017) conclude that it is the efficiency of
gas delivery, whether due to a merger or other process, that sets
the relation between diluted central metallicity and enhanced SFR.
Fig. \ref{doh} also shows that there is no significant dilution for the
rest of the star-forming galaxies, including those elevated above
the main sequence by factors of 2--3.  Apparently, these galaxies
have modestly enhanced their global SFRs either without the influx
of metal-poor gas, or they have already re-enriched their ISM.
Our results are consistent with those of Stott et al. (2014)
who found that galaxies above the main sequence have flatter
abundance gradients, an effect that is most evident for specific
SFRs at least a factor of five in excess of the main sequence
expectation.

Barrera-Ballesteros et al. (2016) have recently found that low
mass galaxies (log (M$_{\star}$/M$_{\odot}$) $<$ 9.2) exhibit
central metallicities that are systematically below the resolved
mass-metallicity relation by $\sim$ 0.1 dex.  Of the 21 galaxies
in the \dsfr\ $>0.5$ bin in Fig. \ref{doh}, six have masses
log (M$_{\star}$/M$_{\odot}$) $<$ 9.2.  Computing the $\Delta$
O/H gradients only for galaxies with total stellar mass above
this threshold does not significantly alter our results,
and the centrally diluted metallicity for galaxies with
\dsfr\ $>$ 0.5 persists.

The compaction model predicts that galaxies with positive \dsfr\
should not only have high central \sigsfr, but also high gas
fractions and surface densities. (e.g. Tacchella et al. 2016a).
High global molecular gas fractions have indeed been measured for
SDSS galaxies above the main sequence (e.g. Saintonge et al.
2012, 2016; Sargent et al. 2014; Violino et al. 2017).  At more
moderate redshifts, galaxies with positive \dsfr\ also appear
to be dustier than their main sequence counterparts (Whitaker et al. 2012).
The gas and dust channeled towards the inner parts of the
galaxy could potentially fuel not only central star formation,
but also a dust-obscured AGN.  Mid-IR selected AGN do indeed
seem to exhibit elevated SFRs both locally and at moderately high 
redshifts (Juneau et al. 2013; Cowley et al.
2016; Ellison et al. 2016; Azadi et al. 2017).

However, to move beyond these indirect elements of support for the
connection between a central excess of gas and the compaction
scenario, it is desirable to directly map, on kpc scales, the distribution
of cold gas as a function of main sequence offset.  Unfortunately,
to achieve this for a significant fraction of MaNGA galaxies is
presently unrealistic observationally.  One possible alternative would be to obtain
such gas and dust maps indirectly.  For example, Brinchmann et al. (2013) present
a technique to use optical emission lines to infer total
gas surface densities.  Applying this technique on a spaxel-by-spaxel
basis to the extant MaNGA data, whereby excesses of gas and
star formation in the same data set could be mapped on kpc scales,
would be of great interest.

\subsection{Negative offsets from the main sequence - where is star formation quenched?}

The topic of galaxy quenching has been extensively discussed
in the literature and several papers have used radial profiles
of the specific SFR or its equivalent to conclude that
quenching occurs from the inside out (e.g. Gonz\'{a}lez Delgado
et al. 2016; Belfiore et al. 2017a; S\'{a}nchez et al. 2017b; Morselli et al. in prep).
In Fig. \ref{ms_dsfr_profiles} we confirm the same trend
in \dsfrs\ in passive galaxies - a radially dependent deficit
of star formation that is largest in the inner few kpc.
However, our results also demonstrate that galaxies below
the main sequence (including the passive population) have
suppressed star formation throughout their disks.  Our results
are consistent with other recent MaNGA investigations which have
found that galaxies with suppressed star formation tend to show
reduced sSFR at all radii (e.g. Belfiore et al. 2017b; Spindler
et al. 2017).

Much of the discussion in the literature concerning the
cessation of star formation has focussed on the link between
the tendency of a galaxy to be passive (or quenched) and its
inner stellar density, either implied through structural
parameters (e.g. Wuyts et al. 2011b; Bluck et al. 2014; Lang et al. 2014;
Omand et al. 2014) or through
measurements of \sigmass\ within the central 1 kpc (e.g. Cheung et al. 2012;
Fang et al. 2013; Barro et al. 2013; Woo et al. 2015; Whitaker et al. 2017). 
Moreover, it is observed that quiescent galaxies have a higher central
\sigmass\ for their M$_{\star}$ (e.g. Fang et al. 2013; Barro et al. 2017)
possibly indicating that galaxies' central mass growth is a pre-cursor
to quenching.  The connection between quiescence and 
central mass concentration, and inside out quenching has also been observed at 
high redshifts, out to $z\sim$ 2 -- 3 (Tacchella et al. 2015a; Barro et al. 
2017; Brennan et al. 2017).  These observations provide complementary evidence
that quenching is linked to mechanisms operating in the central galactic
regions which in turn lead to an inside out shut-down of star formation.

The exact cause of this inside-out quenching is still debated, and could
include the exhaustion of gas in the central regions,  AGN feedback
or through the increasing stability of the gas disk (e.g. Dekel \& Burkert 2014).
The first of these scenarios might be expected to lead to low
gas fractions in galaxies following the starburst phase. However, measurements 
of the molecular gas fraction in post starburst galaxies show that they
can still harbour significant gas reservoirs (French et al. 2015; Rowlands et al.
2015; Suess et al. 2017).
Similarly, galaxy mergers, which can cause boosted central SFRs and represent one
of the mechanisms for `wet compaction', show no depletion in their neutral gas
content (Braine \& Combes 1993; Ellison et al. 2015) and even have elevated
molecular gas fractions (Combes et al. 1994; Violino et al. 2017; Sargent et al.
in prep).  Finally, there is abundant evidence that early type galaxies (which
have little or no on-going star formation) frequently contain significant atomic
and molecular gas reservoirs (Young et al. 2011; 2014; Serra et al. 2012; Davis et al. 2016).
Combined, these observations indicate that \textit{galaxy-wide} gas exhaustion may not be the
primary reason for quenching.  However, measurements of the global gas
content can conclude little about whether the \textit{central} gas
reservoir is depleted.  In the absence of large samples of resolved molecular
gas maps, estimates of gas surface density from dust attenuation (Brinchmann
et al. 2013) may provide some insight.  S\'{a}nchez et al. (2017b) have recently
used this approach to infer depleted gas reservoirs in the centres of
passive galaxies.

An alternative inside-out quenching mechanism that can leave the gas reservoir
largely intact is the increased stability of the gas disk following
the growth of the central stellar bulge.  This process has become known
as `morphological quenching' (Martig et al. 2009).  Indeed, there is
now a wealth of observational evidence that links the prominence
of the bulge/central mass surface density to the quenched fraction
 (e.g. Wuyts et al. 2011b; Bluck et al. 2014; Omand et al. 2014;
Woo et al. 2015).  Morphological
quenching is also supported by observations of lower star formation efficiency
of bulge dominated/early type galaxies (Saintonge et.
al 2012; Martig et al. 2013; Davis et al. 2014).  Gonz\'{a}lez Delgado
et al. (2016) have similarly proposed that morphological quenching
could explain the increasingly suppressed inner sSFR profiles
as a function of morphological type.

However, the link between bulges and quenching has also
been interpreted as possible evidence for AGN driven feedback (e.g. Bluck
et al. 2014; Teimoorinia,
Bluck \& Ellison 2016), a process that might also be expected to operate from
the inside out.  Indeed, simulations whose quenching prescriptions are based on
AGN feedback show very similar trends of passive
fraction with morphology as seen in observations (Bluck et al. 2016; Brennan et al.
2017).  Even more compelling evidence for an AGN-quenching scenario is the tendency for
passive galaxies to host central supermassive black holes that are more massive 
at fixed galactic stellar mass than star-forming galaxies (Savorgnan et al.
2016; Terrazas et al. 2016).

Overall, our results support a model in which galaxies both
boost and quench their star formation from the inside out.
Indeed, there is a remarkable symmetry in the \dsfrs\ profiles
of galaxies far above and below the main sequence (Fig. \ref{ms_dsfr_profiles}).
Our results are
consistent with the framework of the compaction scenario in which
galaxies first experience a centrally concentrated star burst which
builds central mass and is then
followed by depletion/feedback which similarly acts from the centre
outwards (e.g. Tacchella et al. 2016a, b).   At moderately
high redshifts there exists a population of currently star forming
galaxies that are compact, lying on the central \sigmass\ -- M$_{\star}$ relation
for quenched galaxies (e.g. Barro et al. 2013, 2014, 2017), possibly
indicating that they are the pre-cursors to the quenched population.  

\section{Summary}

We have investigated the spatial dependence of enhanced/suppressed star 
formation for galaxies that lie above/below the global star forming
main sequence using IFU observations obtained from the MaNGA survey
(Section 2).
Our sample consists of 392 star-forming galaxies (Fig. \ref{MS}, Section \ref{sf_sec})
whose SFRs extend an order of magnitude above and below the global
main sequence (Fig. \ref{ms_dsfr}, Section \ref{dsfr_sec}).
We additionally include a sample of 470 passive galaxies, defined
as having SFRs more than a factor of ten below the main sequence
(Fig. \ref{MS_passive}, Section \ref{pass_sec}).
The relative star formation in a given \textit{spaxel} (\dsfrs) is
quantified with reference to the resolved main sequence (Fig. 
\ref{manga_sfms}, Section
\ref{res_MS_sec}) by matching to comparison spaxels with similar 
\sigmass\ and radial distance from the galaxy centre (Section \ref{dsfrs_sec}).
We also compute metallicity offsets ($\Delta$ O/H) for spaxels,
relative to the resolved MZR (Fig. \ref{manga_mzr}, Sec. \ref{doh_sec}).

Our principal result is presented in Fig. \ref{ms_dsfr_profiles},
in which we quantify radial profiles of star formation excess/deficit
(\dsfrs) as a function of the galaxy's position on the global main sequence.
Galaxies whose total star formation rates place them at least a
factor of a few above the global main sequence have, on average,
elevated \sigsfr\ out to at least 1.5 $R_e$ ($\sim$ 10 kpc).  However,
\sigsfr\ is particularly enhanced within $\sim$ 0.5 $R_e$ ($\sim$ 3 kpc),
indicating a preferential boost in the star formation in the central regions.
Moreover, galaxies that lie at least a factor of three above the main sequence
exhibit metallicities that are diluted relative to the resolved MZR.
These galaxies have positive $\Delta$ O/H profile gradients that
have central values $\sim -0.2$ dex, and approximately normal metallicities
at 2 $R_e$ (Fig. \ref{doh}).
Passive galaxies mirror the star formation profiles of the galaxies above the
main sequence - their \sigsfr\ profiles are depressed throughout,
with the most significant star formation deficit in the central
0.5 $R_e$.  The \dsfrs\ profile of passive galaxies is morphology
dependent; although all passive galaxies have similar \dsfrs\ profiles
beyond $\sim$ 0.6 $R_e$, galaxies with a significant bulge fraction
(B/T$>$0.5, as measured in the mass) have a factor of two lower \sigsfr\
in their central regions (Fig. \ref{dsfr_bt}).

Taken together, our results add to a growing body of
empirical evidence that star formation activity in galaxies is dominated
by changes in the central regions.  This is consistent with
the emerging model of `galaxy compaction' (e.g. Dekel \& Burkert 2014;
Zolotov 2015; Tacchella et al. 2016a,b), in which galaxies
undergo (one, or a series of) gas inflow events which lead
to a central star burst.  Such `wet compaction' events
can be triggered by a variety of processes including
mergers, bars, disk instabilities or streams (e.g. Scudder et al. 2012;
Ellison et al. 2011).  Although galaxies above the main sequence exhibit
elevated star formation throughout the disk (Nelson et al. 2016; Morselli 
et al. 2017; Magdis et al. 2016) the enhancement is 
greatest in the centre.  In turn, this leads to
inside out mass growth (Nelson et al. 2012, 2016; Wuyts et al. 2012;
P\'{e}rez et al. 2013) and eventual
quenching, either from gas depletion, AGN feedback
(e.g. Teimoorinia et al. 2016; Bluck et al. 2016; Terrazas
et al. 2016), or an increased stability in the disk (Martig et al. 2009). 
The quenching process propogates from the central regions 
outwards (Gonz\'{a}lez Delgado et al. 2016;
Belfiore et al. 2017a; Sanchez et al. 2017b) following these episodes
of compact star formation (Barro et al. 2013, 2014, 2017).

\section*{Acknowledgements}

SLE acknowledges stimulating discussions with, and valuable comments
on the manuscript draft from Franceso Belfiore, Asa Bluck,
Alice Concas, Rosa Gonz\'{a}lez Delgado, Lihwai Lin,
Laura Morselli, Erica Nelson, Dave Patton, Paola Popesso
and the anonymous referee.
BA gratefully acknowledges financial support from the MITACS
Globalink program which funded his participation in this project.
SFS and HIM acknowledge the following grants for their support: 
CONACyt CB-180125, DGAPA-UNAM IA100815 and IA101217. 

Funding for the Sloan Digital Sky Survey IV has been provided by
the Alfred P. Sloan Foundation, the U.S. Department of Energy Office of
Science, and the Participating Institutions. SDSS-IV acknowledges
support and resources from the Center for High-Performance Computing at
the University of Utah. The SDSS web site is www.sdss.org.

SDSS-IV is managed by the Astrophysical Research Consortium for the 
Participating Institutions of the SDSS Collaboration including the 
Brazilian Participation Group, the Carnegie Institution for Science, 
Carnegie Mellon University, the Chilean Participation Group, the French 
Participation Group, Harvard-Smithsonian Center for Astrophysics, 
Instituto de Astrof\'isica de Canarias, The Johns Hopkins University, 
Kavli Institute for the Physics and Mathematics of the Universe (IPMU) / 
University of Tokyo, Lawrence Berkeley National Laboratory, 
Leibniz Institut f\"ur Astrophysik Potsdam (AIP),  
Max-Planck-Institut f\"ur Astronomie (MPIA Heidelberg), 
Max-Planck-Institut f\"ur Astrophysik (MPA Garching), 
Max-Planck-Institut f\"ur Extraterrestrische Physik (MPE), 
National Astronomical Observatories of China, New Mexico State University, 
New York University, University of Notre Dame, 
Observat\'ario Nacional / MCTI, The Ohio State University, 
Pennsylvania State University, Shanghai Astronomical Observatory, 
United Kingdom Participation Group,
Universidad Nacional Aut\'onoma de M\'exico, University of Arizona, 
University of Colorado Boulder, University of Oxford, University of Portsmouth, 
University of Utah, University of Virginia, University of Washington, University of Wisconsin, 
Vanderbilt University, and Yale University.

\appendix

\section{Comparison of MaNGA and DR7 masses and star formation rates}\label{app}

\begin{figure}
	\includegraphics[width=\columnwidth]{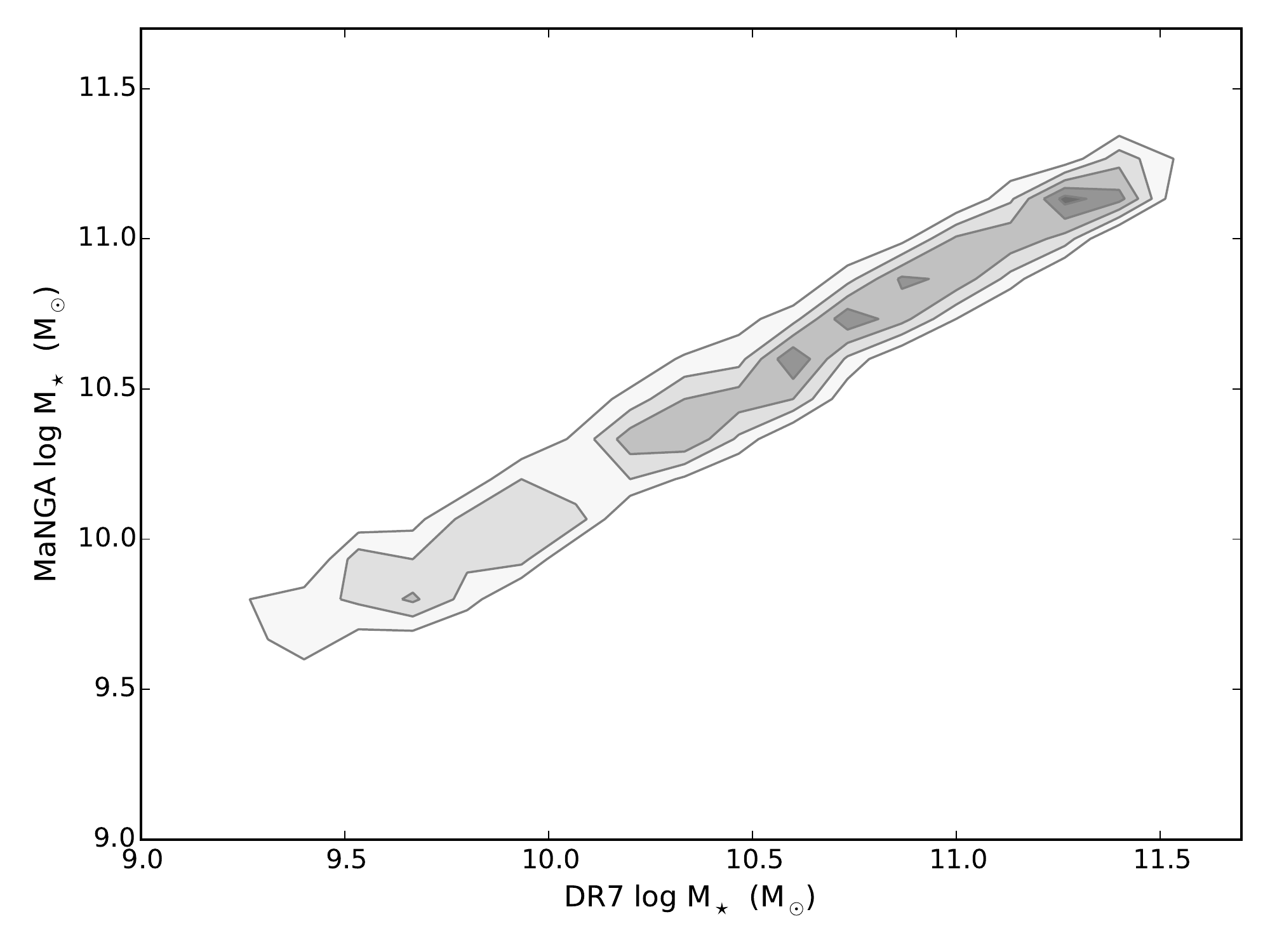}
	\includegraphics[width=\columnwidth]{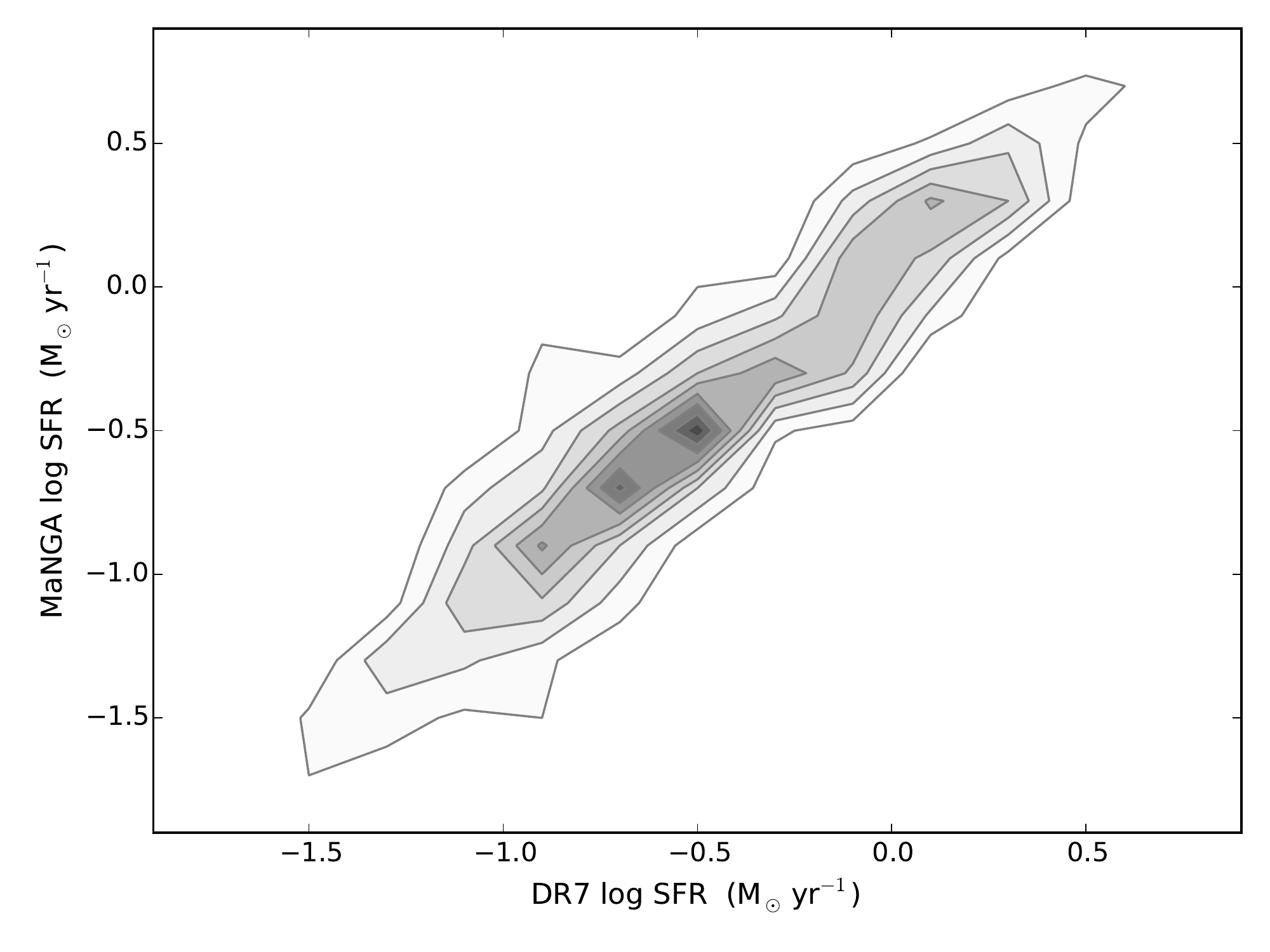}
        \caption{Comparison of stellar masses (top panel) and SFRs (bottom
panel) between MaNGA and the MPA/JHU DR7 catalogs. }
    \label{fig_app}
\end{figure}

In Fig. \ref{fig_app} we compare the integrated (full
IFU) MaNGA star formation rates and stellar masses of all galaxies in the parent 
DR13 sample with the values in the MPA/JHU DR7 catalogs.  Values are corrected for
the different cosmologies and initial mass functions used in the two catalogs. Both the SFRs and stellar
masses trace each other well; as noted in Sec. \ref{sf_sec} the mean difference 
between the MPA/JHU and integrated MaNGA values is 0.0007 dex for stellar mass and 
0.03 dex for SFR, with scatter of $\sim$ 0.3 and 0.4 dex respectively.  Importantly,
we note that any difference between the MaNGA and DR7 values will not affect
our analysis, due to the comparative nature of the methodology that we have
adopted, such that comparisons are always made consistently within a given
sample.

\end{document}